\documentclass[12pt]{article}

\usepackage[utf8x]{inputenc}
\usepackage{amsmath}
\usepackage{amsfonts}

\usepackage{graphicx}
\usepackage[subrefformat=parens,labelformat=parens]{subfig}
\usepackage{color}

\graphicspath{{./Figures/}}

\newcommand{\Gm}{\ensuremath{\text{G}_m}}
\newcommand{\Gma}{\ensuremath{\text{G}_{ma}}}
\newcommand{\Gmi}{\ensuremath{\text{G}_{mi}}}
\newcommand{\Gc}{\ensuremath{\text{G}_c}}
\newcommand{\Gmt}{\ensuremath{\tilde{\text{G}}_m}}
\newcommand{\Gct}{\ensuremath{\tilde{\text{G}}_c}}
\newcommand{\Gmh}{\ensuremath{\hat{\text{G}}_m}}
\newcommand{\Gch}{\ensuremath{\hat{\text{G}}_c}}

\newcommand{\raca}{\ensuremath{\text{Rac}_a}}
\newcommand{\racmi}{\ensuremath{\text{Rac}_{mi}}}
\newcommand{\racci}{\ensuremath{\text{Rac}_{ci}}}
\newcommand{\rhoa}{\ensuremath{\text{Rho}_a}}
\newcommand{\rhomi}{\ensuremath{\text{Rho}_{mi}}}
\newcommand{\rhoci}{\ensuremath{\text{Rho}_{ci}}}
\newcommand{\cdca}{\ensuremath{\text{Cdc42}_a}}
\newcommand{\cdcmi}{\ensuremath{\text{Cdc42}_{mi}}}
\newcommand{\cdcci}{\ensuremath{\text{Cdc42}_{ci}}}
\newcommand{\pone}{\ensuremath{\text{PIP}}}
\newcommand{\ptwo}{\ensuremath{\text{PIP}_{2}}}
\newcommand{\pthree}{\ensuremath{\text{PIP}_{3}}}

\newcommand{\beginsupplement}{%
        \setcounter{table}{0}
        \renewcommand{\thetable}{S\arabic{table}}%
        \setcounter{figure}{0}
        \renewcommand{\thefigure}{S\arabic{figure}}%
        \setcounter{section}{0}
        \renewcommand{\thesection}{S\arabic{section}}%
     }

\begin{document}

\title{Effects of 3D Geometries on Cellular Gradient Sensing and Polarization}
\author{Fabian Spill$^{1,2}$, Vivi Andasari$^{1}$, Michael Mak$^{1,2}$,\\
 Roger D. Kamm$^{2,*}$, Muhammad H. Zaman$^{1,*}$
}

\maketitle
\noindent
$^{1}$Department of Biomedical Engineering, Boston University, 44 Cummington Street, Boston MA 02215, USA.\\
$^{2}$Department of Mechanical Engineering, Massachusetts Institute of Technology, 77 Massachusetts Avenue, Cambridge, MA 02139, USA.\\
$^*$ Co-corresponding authors, rdkamm@mit.edu (RDK), zaman@bu.edu (MHZ)

\begin{abstract}
During cell migration, cells become polarized, change their shape, and move in response to various internal and external cues. Cell polarization is defined through the spatio-temporal organization of molecules such as PI3K or small GTPases, and is determined by intracellular signaling networks. It results in directional forces through actin polymerization and myosin contractions. Many existing mathematical models of cell polarization are formulated in terms of reaction-diffusion systems of interacting molecules, and are often defined in one or two spatial dimensions. In this paper, we introduce a 3D reaction-diffusion model of interacting molecules in a single cell, and find that cell geometry has an important role affecting the capability of a cell to polarize, or change polarization when an external signal changes direction. Our results suggest a geometrical argument why more roundish cells can repolarize more effectively than cells which are elongated along the direction of the original stimulus, and thus enable roundish cells to turn faster, as has been observed in experiments. On the other hand, elongated cells preferentially polarize along their main axis even when a gradient stimulus appears from another direction. Furthermore, our 3D model can accurately capture the effect of binding and unbinding of important regulators of cell polarization to and from the cell membrane. This spatial separation of membrane and cytosol, not possible to capture in 1D or 2D models, leads to marked differences of our model from comparable lower-dimensional models. 
\end{abstract}

\section{Introduction}

The ability to migrate is one of the fundamental properties of cells and is observed in both single-celled organisms as well as multicellular organisms in development, tissue maintenance, and in disease progression. For effective, directional migration, cells need to have the capability to sense and respond to various migratory signals, such as bacteria reacting to nutrients or other attractants or repellents \cite{berg1977physics,wadhams2004making}, cells being guided to their correct location during embryonic development \cite{carlson2013human} or immune cells migrating towards locations of injury or infection \cite{madri2000cell}. Furthermore, cell migration plays a prominent role in diseases such as cancer, where the majority of deaths are caused by metastases. Hence migration, invasion and metastasis are considered defining properties of cancer \cite{hanahan2000hallmarks,hanahan2011hallmarks}.

External stimuli affecting cell migration include biochemical signals \cite{van2004chemotaxis} or mechanical interactions with the environment \cite{lo2000cell,zaman2006migration,borau2011}. One particularly interesting feature is the capability of many cells to detect spatial variations in the concentrations of biochemicals and to migrate towards, or away from the sources of such chemicals. Often, the gradients of those chemoattractants or repellents have a small slope, so cells need a mechanism to detect and magnify external biochemical stimuli \cite{servant2000polarization}. Such gradient detection then enables the cells to develop a polarized state with a well defined front and back. To this purpose, chemical signals need to be translated into the generation of mechanical forces \cite{mogilner2003force}, which ultimately enable the cell to migrate in the direction defined by the polarized state.

In the last few decades, researchers have discovered and studied a large number of key molecules understood to play an important role in the sensing of chemical stimuli as well as the subsequent polarization, regulation of the actin cytoskeleton and generation of mechanical forces \cite{ridley2003cell}. Among these molecules are small GTPases \cite{raftopoulou2004cell,jaffe2005rho}, PI3K, PTEN, phosphatidylinositols (PIPs), \cite{funamoto2002spatial,cully2006beyond,kay2008changing}, Arp2/3 \cite{pollard2007regulation,ryan2012excitable} and Cofilin \cite{yang1998cofilin,arber1998regulation}.

To understand the complexity of those pathways of interacting molecules, as well as to understand the mechanisms of sensing external gradients and polarizing a cell, a large number of mathematical models of gradient sensing and cell polarization have been developed (see \cite{iglesias2008navigating,jilkine2011comparison} for reviews). Whereas some of these mathematical models try to explain the general principles of signal detection, amplification and polarization \cite{Meinhardt1999,levchenko2002models,iglesias2012biased}, others attempt to explicitly model the dynamics and interactions between some of the most important involved molecules \cite{maree2006polarization,dawes2007phosphoinositides,jilkine2007mathematical,otsuji2007mass,
goryachev2008dynamics,holmes2012modelling,rubinstein2012,knoch2014}. Many of these models are formulated in terms of reaction-diffusion partial differential equations (PDEs) and make use of ideas such as pattern formation, which have been applied to biology for many years \cite{turing1952chemical,gierer1972theory,meinhardt1974applications,meinhardt2000pattern}. Alternative modeling approaches to cell polarization include \cite{kozlov2007model}, where thermodynamic considerations were used to predict polarization, \cite{vanderlei2011computational}, where the effects of the interplay of biochemistry and mechanics on polarization were investigated, or \cite{altschuler2008spontaneous}, where stochastic cell polarization was considered. The majority of these mathematical models have been formulated, or at least tested, in one or two spatial dimensions. Simulating a model in those lower dimensions greatly decreases the computational costs, and might seem justified if one is modeling cell migration on 2D substrates or in quasi-one-dimensional scenarios such as the detection of a 1D chemical gradient.

However, when the cell has an irregular shape, it is not a priori clear that a lower dimensional mathematical model can be used. Furthermore, in 3D {\it in vitro} experiments or {\it in vivo}, stimuli can appear from all directions. An additional complication is the spatial organization of the key molecules behind cell polarization and migration: some of the regulators of the actin cytoskeleton, like the Arp2/3 complex, are soluble in the cytosol, whereas others such as phosphatidylinositols, are bound to the membrane. Moreover, some molecules such as the small GTPases can be both membrane bound and soluble, and this binding is influenced by the presence of other regulators such as guanine dissociation inhibitors (GDIs) \cite{dovas2005rhogdi}. Some mathematical models such as \cite{maree2012cells} have studied the influence of cell geometries on cell polarization and migration in two dimensions, the role of cell shape on signaling \cite{meyers2006potential}, the mechanical effects of shape on cell migration \cite{herant2010form}, the effect of cell shape on stress fiber polarization \cite{zemel2010cell}, or the effect of signaling on cell shape \cite{das2012oscillatory,drake2013model}, see also the review \cite{mogilner2009shape}. A mathematical model focusing specifically on the effect of 3D shape on cell polarization, taking into account a whole polarization pathway, has, to our knowledge, not been investigated.

In this paper, we are studying the effect of the cell shape on gradient sensing and cell polarization in a 3D mathematical reaction-diffusion model of key molecules involved in polarization. In section \ref{sec:MembraneCytosolBinding}, we introduce a 3D model of GTPase molecules binding and unbinding from the membrane. We then use this framework to generalize an earlier pathway model investigating the dynamics of the small GTPases Rac, Rho and Cdc42, as well as PIP, PIP2 and PIP3, in a 1D context \cite{holmes2012modelling}, and show how our 3D model can be reduced to such a 1D model. Then, in section \ref{sec:3DSimulations}, we explore how geometry affects the polarization capability and timescales of polarization. We first compare results of the 3D model with simple rectangular geometries to the 1D limit and highlight similarities and key differences. We also find that varying the membrane binding and unbinding rate can change the cell polarization behavior. These rates are altered by the presence of GDI molecules, and while the dynamics of GDI molecules is not included in the present model, our results suggest how GDI molecules will affect cell polarization. We then show how cells with the same volume and length can have vastly different polarization behavior if they have different geometries. Finally, since {\it in vivo} migratory stimuli rarely appear constant in time and space, but dynamically change directions and strength, we investigate how cells react to changes in stimulus, and how this reaction is influenced by geometry. We find that if ellipsoidal cells are initially polarized along their main axis, they cannot adapt to a new stimulus perpendicular to their main axis as efficiently as symmetric, roundish cells. This gives a purely geometrical explanation of the fact that roundish, amoeboid cells can quickly turn and adopt to new stimuli. Furthermore, the ellipsoid cells preferentially polarize along their main axis even if the stimulus gradient is not aligned with this axis. The results in this paper thus predict that cell shape is an important factor influencing the ability of a cell to sense external signals, polarize and ultimately migrate.

\section{Models}\label{sec:MembraneCytosolBinding}
In this section we are introducing a 3D model of cell polarization, and discuss the relation to analogous 1D models. In section \ref{sec:3DcytosolMembraneInteractionModel}, we define a model which consistently describes the binding and unbinding of a molecule to and from the membrane. Then, section \ref{sec:pathway} uses this membrane-cytosol interaction model for inactive GTPases and includes activation of the membrane-bound GTPases, interactions of the three important small GTPases Rho, Rac and Cdc42 as well as interaction with phosphatidylinositols. Finally, in section \ref{sec:oneDimensionalReduction}, we discuss how to reduce our 3D model to a 1D model.
\subsection{3D Membrane-Cytosol Interaction Model}\label{sec:3DcytosolMembraneInteractionModel}
We denote by $\Gc$ the density of molecules which are freely diffusing in the cytosol, measured in moles per volume, and by $\Gm$ the density of membrane-bound molecules, measured in moles per area.  The unbinding rate from the membrane is denoted by $k_{off}$, and the effective binding rate is $k_{on} L_I$, where $k_{on}$ is a conventional rate with dimensions of inverse time, and $L_I$ is the length scale defining the region of the cytosol adjacent to the membrane which is accessible to the membrane-binding reaction. The diffusion coefficients for diffusion in the cytosol or on the membrane, respectively, are denoted $D_C$ and $D_M$. Then, $\Gm$ and $\Gc$ evolve according to the following PDEs:
\begin{align}\label{eq:membraneCytosolBasicEquation}
 \frac{\partial \Gm({\bar r_m},t)}{\partial t} &= D_M \nabla_S^2 \Gm({\bar r_m},t) + k_{on}L_I \Gc({\bar r_m},t) - k_{off}\Gm({\bar r_m},t)\nonumber\\
\frac{\partial \Gc({\bar r_c},t)}{\partial t} &= D_C \nabla_V^2\Gc({\bar r_c},t) \nonumber\\
 -D_C e_n \nabla_V \Gc({\bar r_m},t) &= k_{on}L_I \Gc({\bar r_m},t) - k_{off}\Gm({\bar r_m},t).
\end{align}
The boundary condition for $\Gc$ ensures conservation of the number of molecules under binding and unbinding, and $e_n$ is the unit outward normal vector at the membrane, so $e_n \nabla_V$ is the projection of the gradient on the normal vector. ${\bar r_m}$ and ${\bar r_c}$ denote points on the membrane or in the cytosol, respectively, and $\nabla_V^2$, $\nabla_S^2$ denote the volume and surface Laplace operators (otherwise known as Laplace-Beltrami operator, or Laplacian), respectively. Similar models as \eqref{eq:membraneCytosolBasicEquation} have been used in \cite{madzvamuse2015stability} in the context of diffusion-driven instabilities. The boundary condition is also similar to the boundary conditions chosen to model the flux through a membrane, as done, for instance, in \cite{eliavs2014dynamics}. Such boundary conditions are known as Kedem–Katchalsky boundary conditions. A more detailed discussion and derivation of those equations is provided in the section \ref{sec:App:MembraneCytosolBinding} of the supplementary information. We note that from \eqref{eq:membraneCytosolBasicEquation}, it follows that when $\Gm$ and $\Gc$ are in equilibrium, and are homogeneously distributed, then the fraction $f$ of membrane-bound molecules is given by
\begin{equation}\label{eq:fractionBoundGeneral}
 f = \frac{k_{on}}{k_{on}+k_{off}\frac{V}{L_I S}},
\end{equation}
where $V$ is the volume and $S$ the surface area of the cell.


\subsection{Pathway Model}\label{sec:pathway}
\begin{figure}[!htbp]
  \includegraphics[width=0.95\linewidth]{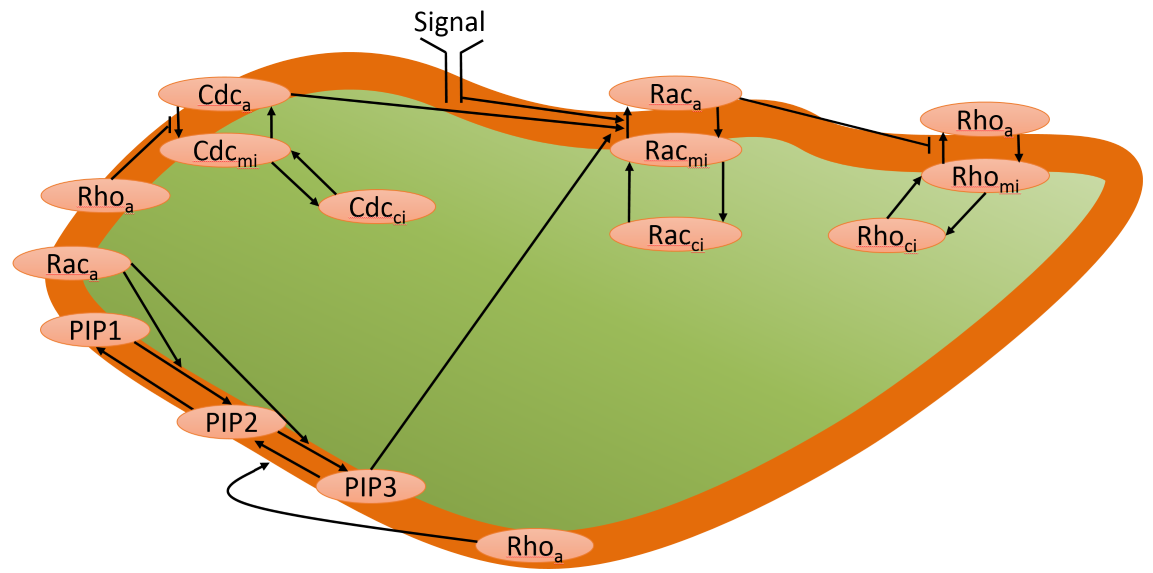}
  \caption{\label{fig:pathway} The molecular pathway considered in this paper incorporates the GTPases Rac, Rho and Cdc42, and the phosphatidylinositols PIP, PIP2 and PIP3. The subscripts, a, mi and ci denote active, membrane-bound inactive and cytosolic inactive GTPases, whereas the PIPs are all membrane-bound. We consider the interactions of active GTPases and PIPs, the activation and deactivation of membrane-bound GTPases, and the binding and unbinding from the membrane of the inactive GTPases.
}
\end{figure}

We now focus on a 3D cell polarization model incorporating the three GTPases Rac, Rho and Cdc42, as well as the three phosphatidylinositols PIP1, PIP2 and PIP3, as dynamic quantities. We use the molecular interactions as shown in Fig. \ref{fig:pathway}, which were previously considered in a 1D model \cite{holmes2012modelling}. PIP, PIP2 and PIP3 are all assumed to be purely membrane bound, whereas the three GTPases Rho, Rac and Cdc42 are assumed to exist in active and inactive membrane-bound forms, indicated by subscripts $a$ and $mi$, respectively, as well as an inactive soluble form, which can diffuse in the cytosol and which is indicated by a subscript $ci$. The membrane binding and unbinding of the inactive forms is described as outlined in section \ref{sec:3DcytosolMembraneInteractionModel}. The full model is thus described by equations \eqref{eq:3DmodelEquations}.
\begin{eqnarray}
\label{eq:3DmodelEquations}
\frac{\partial \raca}{\partial t} & = & D_{M} \nabla^2_S \raca - \delta_{R} \raca + I_{Rac} \frac{\racmi}{\text{Rac}_{tot,2}}
\nonumber 
\\ 
\frac{\partial \racmi}{\partial t} & = & D_M \nabla^2_S\racmi + \delta_{R} \raca - I_{Rac} \frac{\racmi}{\text{Rac}_{tot,2}} + k_{on}L_I\racci - k_{off}\racmi \nonumber\\
\frac{\partial \racci}{\partial t} & = & D_C \nabla^2_V \racci \nonumber\\
 -D_C e_n \nabla_V \racci &=& k_{on}L_I\racci - k_{off}\racmi \nonumber\\
\frac{\partial \rhoa}{\partial t} & = & D_{M} \nabla^2_S \rhoa - \delta_{\rho} \rhoa + I_{\text{Rho}} \frac{\rhomi}{\text{Rho}_{tot,2}}
\nonumber 
\\ 
\frac{\partial \rhomi}{\partial t} & = & D_M \nabla^2_S \rhomi + \delta_{\rho} \rhoa - I_{\text{Rho}}\frac{\rhomi}{\text{Rho}_{tot,2}}+k_{on}L_I\rhoci - k_{off}\rhomi \nonumber \\ 
\frac{\partial \rhoci}{\partial t} & = & D_C \nabla^2_V \rhoci \nonumber\\
 -D_C e_n \nabla_V \rhoci &= & k_{on}L_I\rhoci - k_{off}\rhomi \nonumber \\ 
\frac{\partial \cdca}{\partial t} & = & D_{M} \nabla^2_S \cdca - \delta_{C} \cdca + I_{\text{Cdc}}\frac{\cdcmi}{\text{Cdc}_{tot,2}} 
\nonumber
\\ 
\frac{\partial \cdcmi}{\partial t} & = & D_M \nabla^2_S \cdcmi + \delta_{C} \cdca - I_{\text{Cdc}}\frac{\cdcmi}{\text{Cdc}_{tot,2}}+\nonumber\\
 &+&k_{on}L_I\cdcci - k_{off}\cdcmi
\nonumber
\\ 
\frac{\partial \cdcci}{\partial t} & = &D_C \nabla^2_V \cdcci \nonumber\\
 -D_C e_n \nabla_V \cdcci &=& k_{on}L_I\cdcci - k_{off}\cdcmi
\nonumber
\\ 
\frac{\partial \pone}{\partial t} & =& D_{P} \nabla^2_S \pone - \delta_{P_1} \pone + I_{P_1} + k_{21}\ptwo \nonumber\\
&-& \frac{k_{\text{PI5K}}}{2} \left( 1 + \frac{\raca}{\text{Rac}_{tot,2}} \right) \pone
\nonumber
\\ 
\frac{\partial \ptwo}{\partial t} & = &D_{P} \nabla^2_S \ptwo - k_{21}\ptwo + \frac{k_{\text{PI5K}}}{2} \left( 1 + \frac{\raca}{\text{Rac}_{tot,2}}\right) \pone
\nonumber\\
& -& \frac{k_{\text{PI3K}}}{2} \left( 1 + \frac{\raca}{\text{Rac}_{tot,2}} \right) \ptwo + \frac{k_{\text{PTEN}}}{2} \left( 1 + \frac{\rhoa }{\text{Rho}_{tot,2}}\right) \pthree 
\nonumber
\\ 
\frac{\partial \pthree}{\partial t} & = &D_{P} \nabla^2_S \pthree + \frac{k_{\text{PI3K}}}{2} \left( 1 + \frac{\raca}{\text{Rac}_{tot,2}} \right) \ptwo \nonumber\\
&-& \frac{k_{\text{PTEN}}}{2} \left( 1 + \frac{\rhoa}{\text{Rho}_{tot,2}}\right) \pthree
\end{eqnarray}

Here, the activation functions $I_G$ for the three GTPases are given by
\begin{align}\label{eq:activationFunctions}
I_{Rac} &= \left(I_{R1} + I_{R2}f_1 \frac{PIP_3}{P_{3b}} + \alpha \cdca + S_{Rac}({\bar r_m},t)\right),\nonumber\\
I_{\text{Rho}} &= \frac{I_{\text{Rho}}}{1+\left( \dfrac{\raca}{a_2} \right)^{n}} , \nonumber\\
I_{\text{Cdc}} &= \frac{I_{\text{Cdc}}}{1+\left( \dfrac{\rhoa}{a_1} \right)^{n}} ,
\end{align}
and $\delta_G$ are the deactivation rates. The signal $S_{Rac}({\bar r_m},t)$ is defined to simulate the effect of membrane receptor stimulation of Rac, i.e. it increases the Rac activation rate in a spatial way, and is thus defined on points ${\bar r_m}$ on the membrane. Typically, we will choose a function monotonously increasing along the direction of an external growth factor stimulus. More details about the relation of this model to the 1D model of \cite{holmes2012modelling} are found in the supplementary information, section \ref{sec:App:3DSimulations}. In Table \ref{tab:Parameters}, we also include a full list of the parameters appearing in our model defined by equations \eqref{eq:3DmodelEquations} and \eqref{eq:activationFunctions}.

\subsection{1D Reduction}\label{sec:oneDimensionalReduction}
We now consider the reduction of equation \eqref{eq:membraneCytosolBasicEquation} to a cylindrical cell of length $L$ and radius $R$, where we assume cylindrical symmetry and no strong spatial dependence in the radial direction of the cylinder. Then, equation \eqref{eq:membraneCytosolBasicEquation} reduces to
\begin{align}\label{eq:cylinderReductionTo1D}
 \frac{\partial{\Gmt}(z,t)}{\partial t} &= D_M\partial_z^2 {\Gmt}(z,t) + 2k_{on}\frac{L_I}{R} {\Gct(z,t)} - k_{off}{\Gmt}(z,t),\nonumber\\
 \frac{\partial {\Gct}(z,t)}{\partial t} &= D_C\partial_z^2 {\Gct}(z,t) + \left( k_{off}{\Gmt}(z,t)-2k_{on}\frac{L_I}{R} {\Gct}(z,t) \right).
\end{align}
Here, ${\Gmt}$ and ${\Gct}$ are the densities in one spatial dimension obtained from reducing $\Gm$ and $\Gc$ by ${\Gmt}(z,t) \approx 2\pi R \Gm(\phi,z,t)$, ${\Gct}(z,t) \approx \pi R^2 \Gc(r,\phi,z,t)$, using cylindrical coordinates with radius $r$, angle $\phi$ and axis $z$. In the derivation, we have made use of the assumptions
\begin{align}\label{eq:condition1Dmodel1}
\frac{RL_I{k_{on}}}{3} \ll D_C, \frac{k_{off}R^2}{6}\ll D_C, 
\end{align}
which indicate that radial diffusion is fast, so concentrations equilibrate fast in the radial direction. Note that while equations \eqref{eq:cylinderReductionTo1D} are defined on a 1D spatial domain defined by the length of the cylinder, so $0\leq z \leq L$, the cylinder radius is implicitly present in the sense that the membrane-binding rate $k_{on}$ is effectively renormalized by the inverse of the cylinder radius $R$. If we consider a cell with a given volume $V=\pi R^2 L$, then, while maintaining a cylindrical shape, increasing the length $L$ of the cylinder will result in a decrease of the radius  $R$. Hence, the effective membrane binding rate $2k_{on}\frac{L_I}{R}$ will increase. This makes intuitive sense as for a longer and thinner cylinder, proportionally more molecules in the cytosol are close to the membrane. Indeed, the thin layer of width $L_I$ around the membrane, which is the region of the cytosol accessible to membrane-binding of the molecules, becomes larger for smaller $R$ under fixed cylinder volume. The fraction $f$ of membrane-bound molecules obtained for the 1D cell is given by
\begin{align}\label{eq:fractionBound1Dcylinder}
f = \frac{k_{on}}{k_{on}+k_{off}\sqrt{\frac{V}{4\pi L_I^2L}}},
\end{align}
which differs from the result of \cite{holmes2012modelling}. More details of the derivation of the 1D limit are presented in the supplementary information, section \ref{sec:App:oneDimensionalReduction}, and section \ref{sec:App:twoDimensionalReduction} gives an analogous derivation for the reduction to two spatial dimensions.


%
%
%
%
%
\section{Results}\label{sec:3DSimulations}
In section \ref{sec:3DSimulations1Dlimit}, we consider quasi-one-dimensional cells in our 3D framework and compare this with established 1D models. Then, in section \ref{sec:parameterStudy}, we will investigate the role of the membrane unbinding rate on cell polarization. Finally, in section \ref{sec:3DSimulationsGeometry} we will investigate how 3D geometry can influence the capability and timescales of cells to polarize and to repolarize when the external signal is changing directions.

%

\subsection{Polarization of a Quasi-One-Dimensional Cell}\label{sec:3DSimulations1Dlimit}
\begin{figure}
 \subfloat[$t=0s$]
{
 \includegraphics[width=0.2\linewidth]{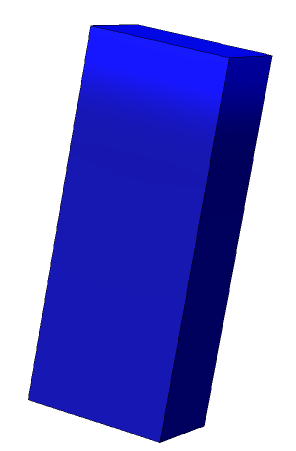}
}
\subfloat[$t=20s$]
{
 \includegraphics[width=0.2\linewidth]{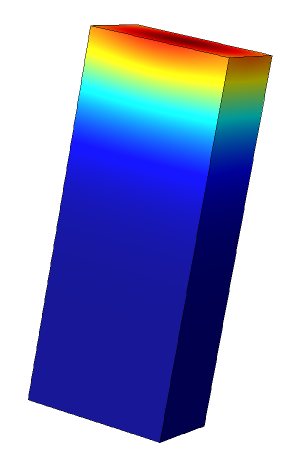}
}
\subfloat[$t=40s$]
{
 \includegraphics[width=0.2\linewidth]{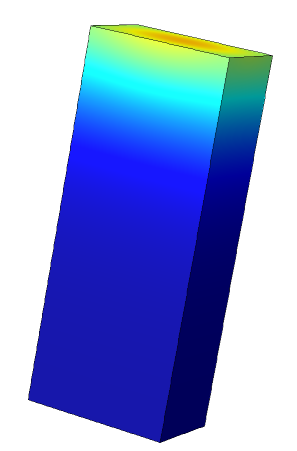}
}
\subfloat[$t=200s$]
{\label{fig:NonPolarizedRectangular200sec}
 \includegraphics[width=0.2\linewidth]{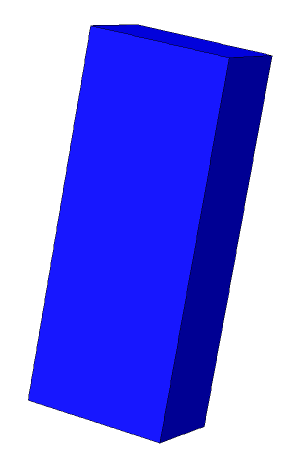}
}
 \includegraphics[width=0.13\linewidth]{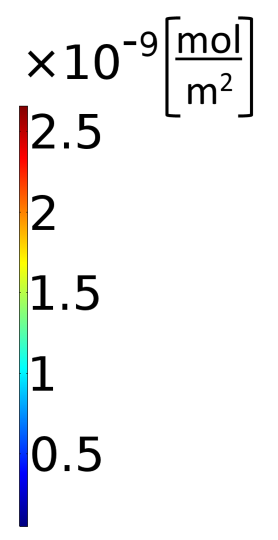}
 \caption{\label{fig:NonPolarizedRectangular} Transient polarization. We show the concentration of active Cdc42 on the membrane for a rectangular cell with side lengths $L,w,d=20,8,5\mu m$. Active Rac is perturbed at the top of the cell, leading to a brief polarized state which then fades away with progressing time.}
\end{figure}

To compare to the 1D model \cite{holmes2012modelling}, we are now investigating a scenario of emerging polarization where we start with initially homogeneous concentrations of all molecules, which are then perturbed by a large spike in active Rac at one end of the long cell in a symmetric way depending only on the direction of the longest extent of the cell. First of all, we checked our code on a cuboid-shaped cell with side lengths $L,w,d=20,8,5 \mu m$, since a cuboid presents the most straight-forward generalization of a 1D geometry. Fig. \ref{fig:NonPolarizedRectangular} shows a typical time evolution of a GTPase concentration, here Cdc42, after the initial Rac stimulus is applied at time $t=0$ at the top of the cell, which is then removed. We observe that at time $t=20s$ it looks as if the cell could polarize, but the strength of polarization fades away and is completely absent at time $t=200s$.
As we are interested in studying the effect of geometry in this paper, we vary the length of the cell, fixing the cell volume to $V=800\mu m^3$. 
\begin{figure}
\begin{center}
\subfloat[$L=40\mu m$, $t=200s$]
{
 \includegraphics[width=0.3\linewidth]{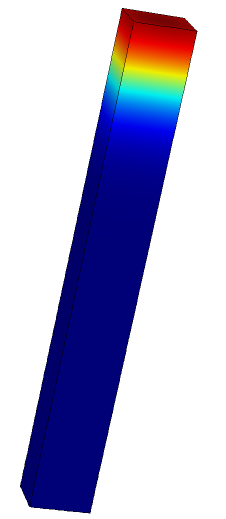}
}
\subfloat[$L=80\mu m$, $t=200s$]
{
 \includegraphics[width=0.3\linewidth]{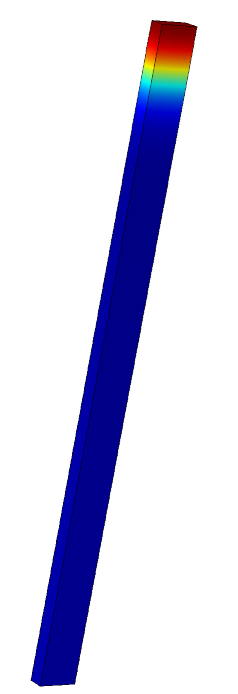}
}
 \includegraphics[width=0.2\linewidth]{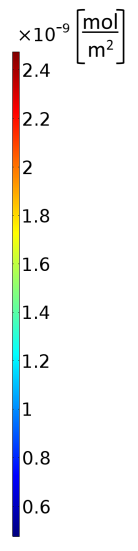}
 \end{center}
\caption{\label{fig:PolarizationOfRectangularDifferentLength} Persistent polarization. The same cell and setup as in Fig. \ref{fig:NonPolarizedRectangular}, but elongated to $L=40$ and $80\mu m$. We see that, contrary to the cell with $L=20\mu m$ shown in Fig. \protect\subref*{fig:NonPolarizedRectangular200sec}, after $200 s$ a stable polarized state is maintained.
}
\end{figure}

Fig. \ref{fig:PolarizationOfRectangularDifferentLength} shows this cell with different lengths, $L=40, 80 \mu m$. In each case we perturbed active Rac at the top of the cell and show the active Cdc42 concentrations after $t=200s$. We see that in both cases a stable polarization pattern is established. This is in contrast to the case of the shorter cell with $L=20\mu m$, which, as shown in Fig. \ref{fig:NonPolarizedRectangular}, has no signs of polarization after $t=200s$. Fig. \ref{fig:App:PolarizationOfRectangularDifferentLengthInTime} in the supplementary information shows results from the same simulations but focuses on the time series of active Cdc42 at the front and back of the cell. These results are compatible with the observation in \cite{holmes2012modelling} that length can change the bifurcation behavior and increase the polarization sensitivity. However, our results are different for several reasons: we take the finiteness of the membrane binding and unbinding rates into account; the fraction of membrane-bound GTPases which we derived in the section \ref{sec:oneDimensionalReduction} is different from the one used in \cite{holmes2012modelling}; \cite{holmes2012modelling} combined the inactive membrane-bound and cytosolic forms into one inactive form whereas we do not perform this approximation in the 3D model; \cite{holmes2012modelling} measured the membrane-bound particles in moles per volume, whereas we use moles per unit area, which is important as we fix the volume, but by changing the length also change the surface area of the cells.

\begin{figure}
\subfloat[Homogeneous]
{\label{fig:Polarization1Dversus3DHomogeneous}
 \includegraphics[width=0.48\linewidth]{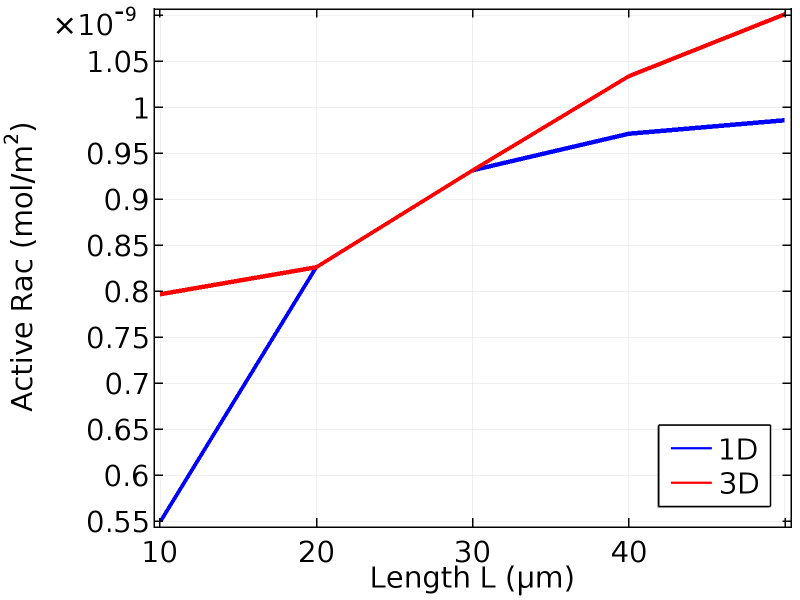}
}
\subfloat[Gradient]
{\label{fig:Polarization1Dversus3DGradient}
 \includegraphics[width=0.48\linewidth]{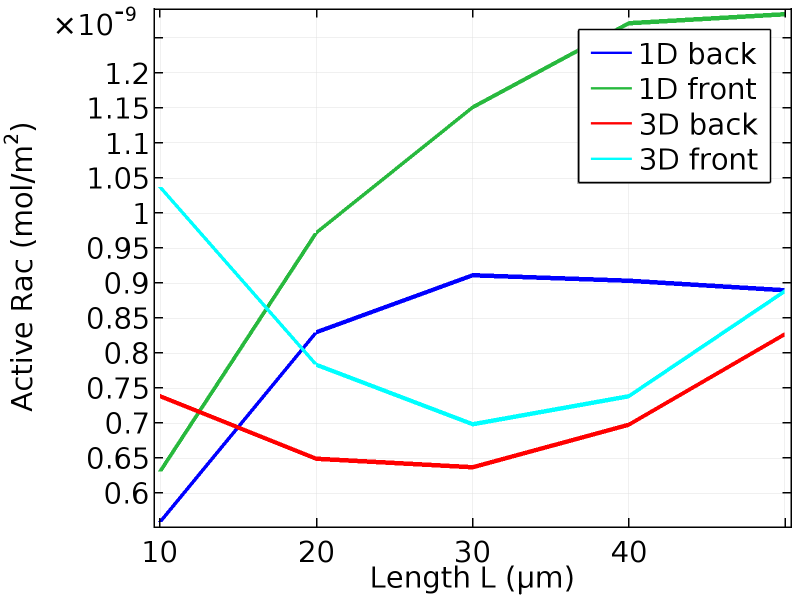}
}
\caption{\label{fig:Polarization1Dversus3D} Length dependence of active Rac in homogeneous conditions \protect\subref{fig:Polarization1Dversus3DHomogeneous} and in the presence of a gradient \protect\subref{fig:Polarization1Dversus3DGradient}, in which case the concentrations at both the front and back are shown. In each case we compare the 1D model from \cite{holmes2012modelling} with our 3D model, and the parameters are aligned so that, with homogeneous conditions and $L=20\mu m$, both models coincide.
}
\end{figure}
In Fig. \subref*{fig:Polarization1Dversus3DHomogeneous}, we show active Rac levels in a homogeneous, steady state setting, and in Fig. \subref*{fig:Polarization1Dversus3DGradient}, active Rac is shown in the presence of a constant linear gradient stimulus $200 s$ after this stimulus is initially applied, for different lengths of the cell in the 1D and 3D models. As before, the volume of the cell is fixed at $V=800\mu m^3$. We have chosen the parameters of the 3D model such that at the base length of $L=20\mu  m$, we get agreement with the 1D model and homogeneous conditions. Fig. \subref*{fig:Polarization1Dversus3DHomogeneous} shows that the steady state values obtained in the homogeneous case differ significantly when the length is changed. Moreover, when the gradient is applied, the results presented in Fig. \subref*{fig:Polarization1Dversus3DGradient} confirm that the differences in Rac active concentrations at the front and back of the cell can differ markedly between the 1D and 3D model. In the scenario shown, in the 3D model, the difference of active Rac at the front and the back decreases with increasing length, whereas in the 1D model it increases with increasing length.

\begin{figure}
\subfloat[]
{\label{fig:asymmetryDependence}
 \includegraphics[width=0.48\linewidth]{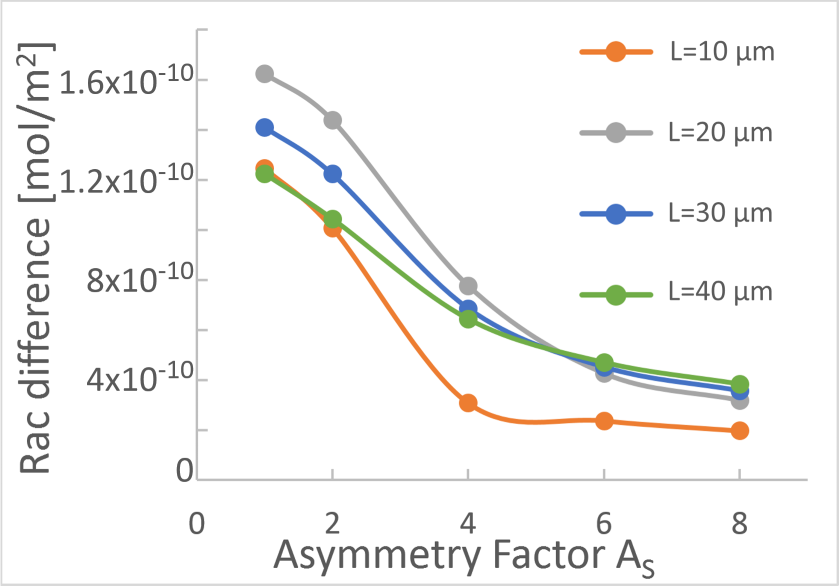}
}
\subfloat[]
{\label{fig:volumeDependence}
 \includegraphics[width=0.48\linewidth]{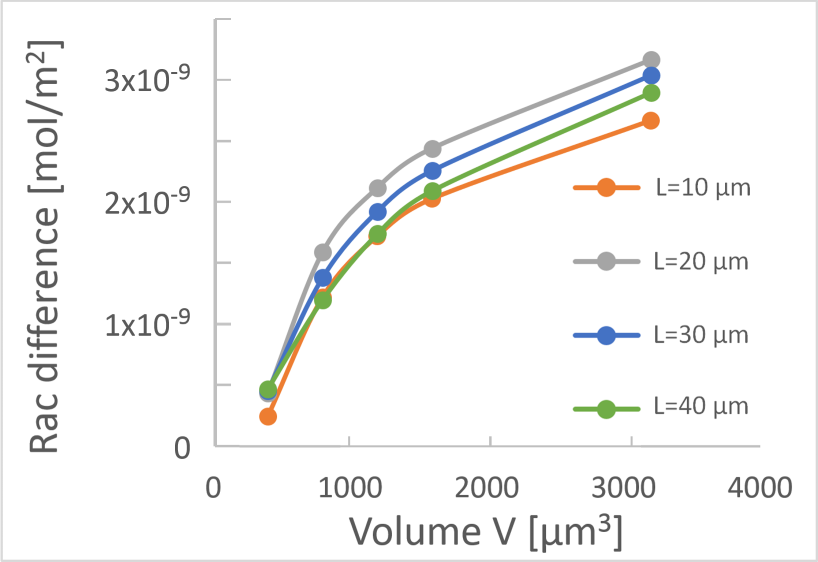}
}
\caption{
\label{fig:AsymmetryAndVolumeDependence} Geometry influences Polarization strength. As a measure of polarization strength, we show the difference of active Rac on the membrane at the front and at the back of the cell, for cells of different rectangular shapes. Each plot shows four graphs for cells of different lengths, $L=10 \mu m$ (orange), $20 \mu m$ (gray), $30\mu m$ (blue), $40\mu m$ (green). In (a) the volume is fixed to be $V=800\mu m^3$, and the sides $w\geq d$ perpendicular to $L$ of the rectangle are fixed in dependence on the asymmetry factor $A_S$ as $w=\sqrt{\frac{A_S V}{ L}}$, $d=\sqrt{\frac{V }{A_S L}}$. In (b), the volume dependence is shown while $w=d=\sqrt{\frac{V}{L}}$.
}
\end{figure}
Furthermore, in a 1D model one cannot accurately take into account variations in the directions perpendicular to the main axis. In Fig. \subref*{fig:asymmetryDependence}, we investigate the impact of the asymmetry factor $A_S = \frac{w}{d}$, which describes the asymmetry of the directions perpendicular to the length $L$, on polarization. We find that, for cells of different length, higher asymmetry decreases polarization strength, measured in terms of the difference of active Rac between the front and the back of the cell. Then, in Fig. \subref*{fig:volumeDependence}, we investigate the volume dependence, and find that generally, increasing the volume $V$ of the cell increases polarization strength for cells of different lengths. A main effect of changing either volume or asymmetry is that this will change the volume to surface ratio, which then affects effective activation and inactivation rates as well as effective membrane binding and unbinding rates. Furthermore, the effective diffusion rates are changed when volume or asymmetry change.

\subsection{Role of Membrane Unbinding Rates}\label{sec:parameterStudy}
We now investigate the dependence of our model on an important new parameter typically not considered in previous models, that is, the membrane unbinding rate $k_{off}$. Its associated binding rate $k_{on}$ is fixed via relation \eqref{eq:fractionBoundGeneral}. It is of physiological importance, as GDI molecules mediate the sequestration of GTPases into the cytosol \cite{garcia2011invisible}, and hence their dysregulation will change binding and unbinding rates. Hence, it is important to know how the model predictions change when these rates are varied.
\begin{figure}
\subfloat[$L=20\mu m$]
{\label{fig:offRateDependenceLength20}
 \includegraphics[width=0.48\linewidth]{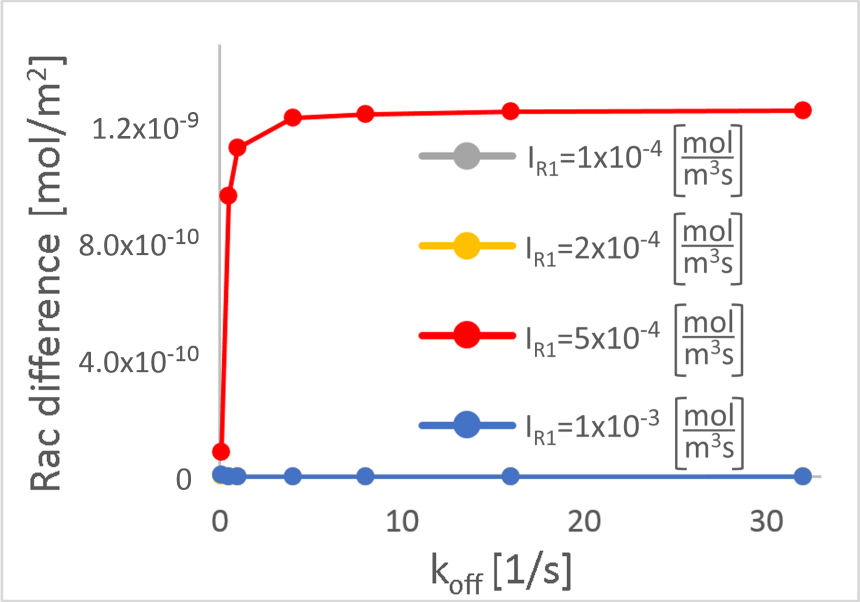}
}
\subfloat[$L=40\mu m$]
{\label{fig:offRateDependenceLength40}
 \includegraphics[width=0.48\linewidth]{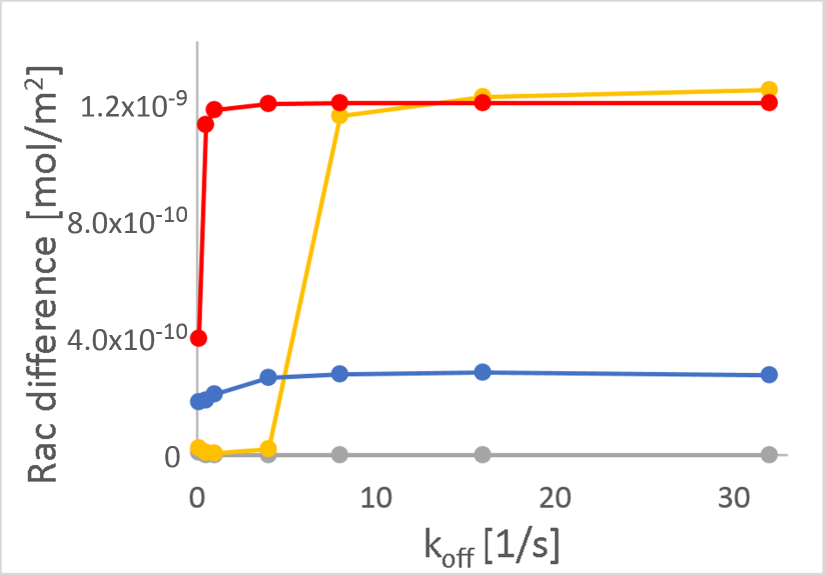}
}
\caption{
\label{fig:offRateDependence} Membrane unbinding influences polarization strength. 
As a measure of polarization strength, we compare the difference in concentrations of active Rac between the front and the back of the cell as a function of $k_{off}$ for different values of the Rac activation rate $I_{R1}$ and two lengths, \protect\subref{fig:offRateDependenceLength20} $L=20\mu m$ and \protect\subref{fig:offRateDependenceLength40} $L=40\mu m$, in all cases 200 seconds after an initial stimulus of active Rac is applied at the front of the cell. In \protect\subref{fig:offRateDependenceLength20}, only the cell with $I_{R1}=0.0005$ shown in red is polarized, and all three other values lead to completely unpolarized states.
}
\end{figure}
Fig. \ref{fig:offRateDependence} shows the difference of active Rac between the front and back, which is a measure of the polarization strength, of a rectangular cell as a function of $k_{off}$ $200 s$ after an initial stimulus of Rac at the front. We show this dependence for different cell lengths and values of the Rac activation rate $I_{R1}$, as these parameters where shown in \cite{holmes2012modelling} to be important parameters affecting cell polarization. Fig. \subref*{fig:offRateDependenceLength20} shows results from a cell of length $L=20\mu m$, whereas Fig. \subref*{fig:offRateDependenceLength40} shows results from a cell of $L=40\mu m$. In each case, we show plots for four different Rac activation rates $I_{R1}$. First, we confirm qualitatively the observation of \cite{holmes2012modelling} that intermediate ranges of $I_{R1}$ can lead to a polarized states, or are more strongly polarized. Furthermore, the shorter cell does not polarize as easily as the longer cell, as in Fig. \subref*{fig:offRateDependenceLength20} only the cell with $I_{R1}=0.0005$ is polarized. We also see that $k_{off}$ is positively associated with polarization strength, and the cells with very small values of $k_{off}=0.1s^{-1}$ do not, or only weakly, polarize. Note that the 1D limiting case requires $k_{off}\gg \frac{D_C}{L^2} = \frac{1}{4}s^{-1},\frac{1}{16} s^{-1}$ for $L=20,40\mu m$, respectively, so in neither case are the approximations applied in \cite{holmes2012modelling} necessarily expected to be accurate. This constraint is discussed along equation \eqref{eq:condition1Dmodel2} in the supplementary information. We also note that for most, but not all parameters checked, the polarization strength saturates at $k_{off}$ values of the order of magnitude of $1 s^{-1}$. The results in Fig. \ref{fig:offRateDependence} confirm that the membrane unbinding rate is an important parameter which can influence the capability of a cell to polarize.

\subsection{Influence of Geometry on Polarization}\label{sec:3DSimulationsGeometry}

We now investigate how cell shape influences the ability of the cell to polarize, lose polarization or repolarize when the direction of a signal changes in time. Here we include some effects which cannot be investigated with a 1D model.

\subsubsection{Influence of Geometry on Initial Polarization}
\begin{figure}[h]
\centering
\subfloat[]
{
\label{fig:RealTransmigratingCell1}
 \includegraphics[width=0.22\linewidth]{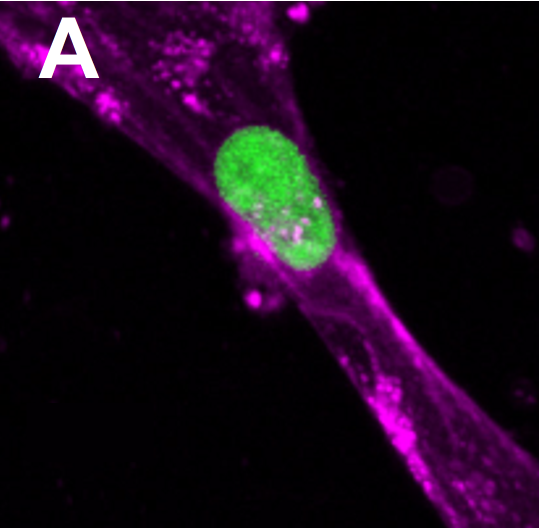}
}
\subfloat[]
{
\label{fig:RealTransmigratingCell2}
 \includegraphics[width=0.22\linewidth]{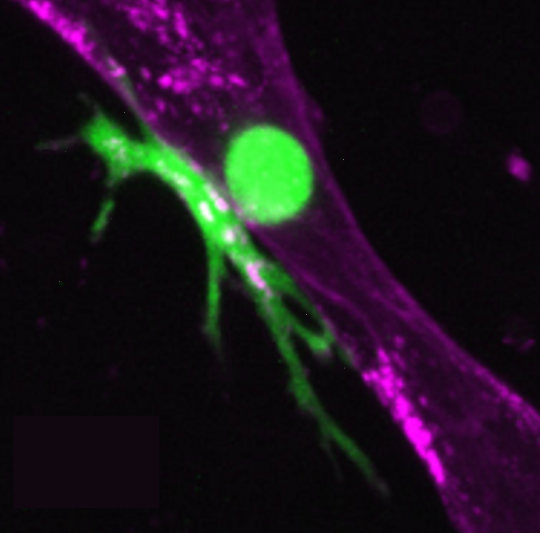}
}
\caption{\label{fig:RealCells}
Cells can appear in vastly different shapes. Here, a cancer cell
(green) extravastating from inside a vessel of endothelial cells (purple) into surrounding extracelullar matrix (black) is shown, as observed in \cite{chen2013mechanisms} (Reproduced by permission of The Royal Society of Chemistry). In (a) the cancer cell appears nearly spherical, while
it is still fully inside the vessel lumen and has not started to extravasate. When it is
in the process of extravasation through the endothelium, it narrows dramatically at the endothelium, connected only through a thin neck region (b). Part of the cell remains in the lumen, but much of it has already spread outside of the lumen into the extracellular matrix.
}
\end{figure}
\begin{figure}[h]
 \includegraphics[width=0.2\linewidth]{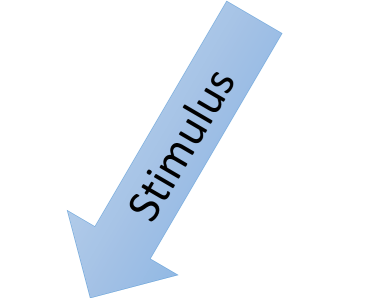}
\subfloat[t=5s]
{\label{fig:PolarizationDependingOnShapeSphere2}
 \includegraphics[width=0.2\linewidth]{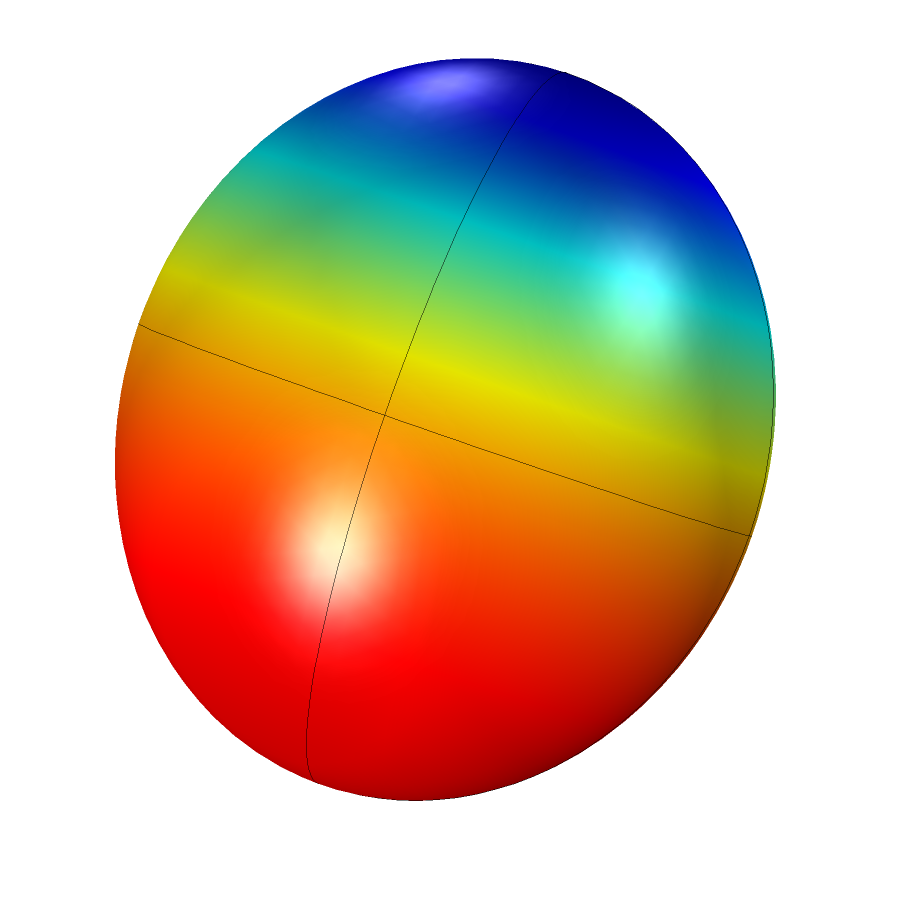}
}
\subfloat[t=10s]
{\label{fig:PolarizationDependingOnShapeSphere3}
 \includegraphics[width=0.2\linewidth]{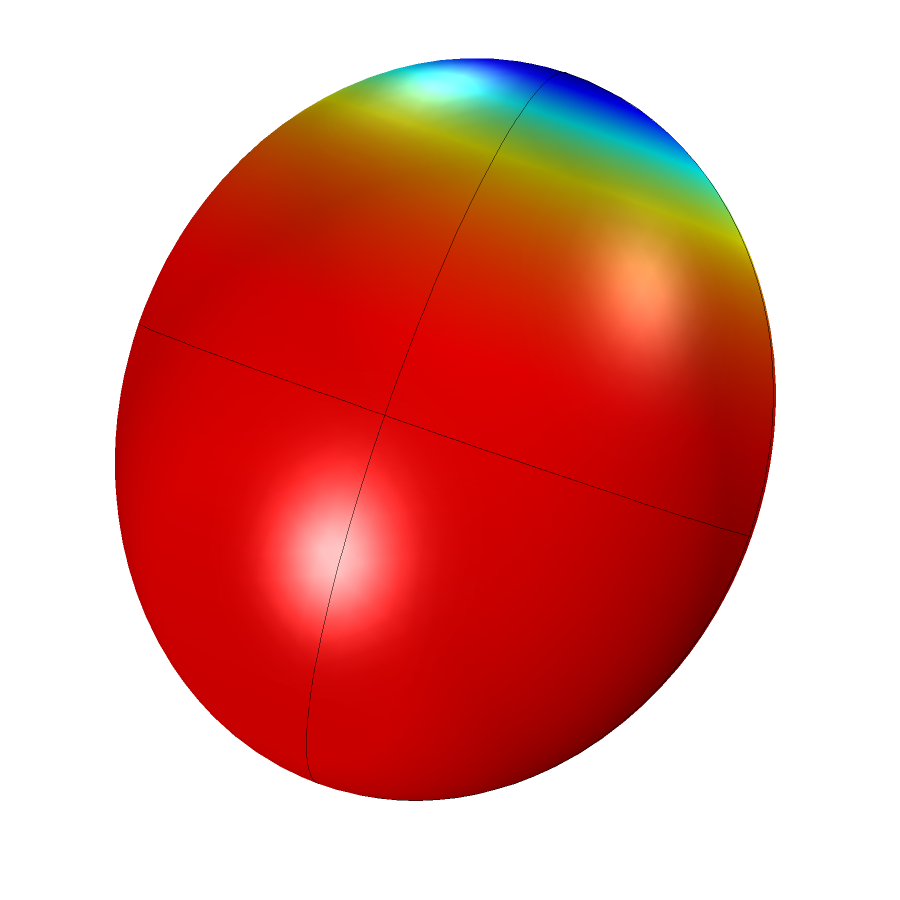}
}
\subfloat[t=100s]
{\label{fig:PolarizationDependingOnShapeSphere4}
 \includegraphics[width=0.2\linewidth]{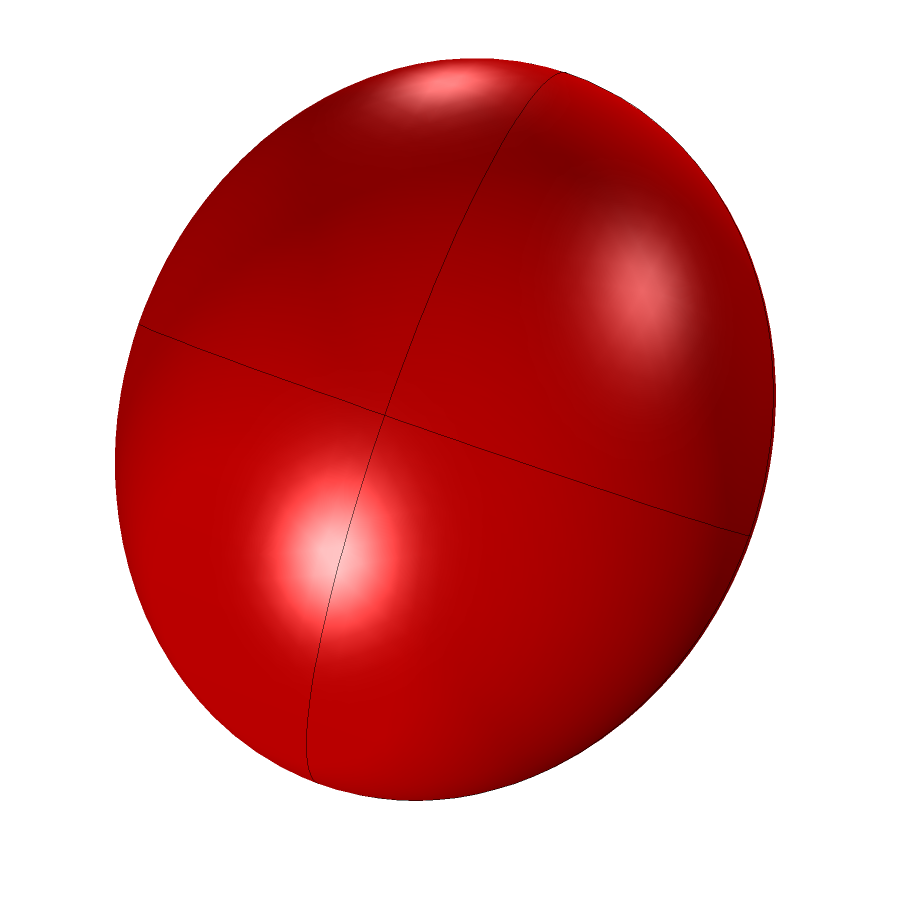}
}
\includegraphics[width=0.13\linewidth]{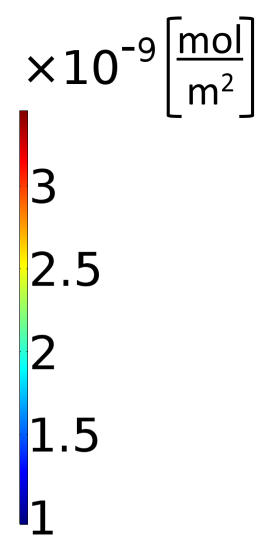}
\\
\includegraphics[width=0.2\linewidth]{Stimulus_arrow_extravasation}
\subfloat[t=5s]
{\label{fig:PolarizationDependingOnShapeEllips2}
 \includegraphics[width=0.2\linewidth]{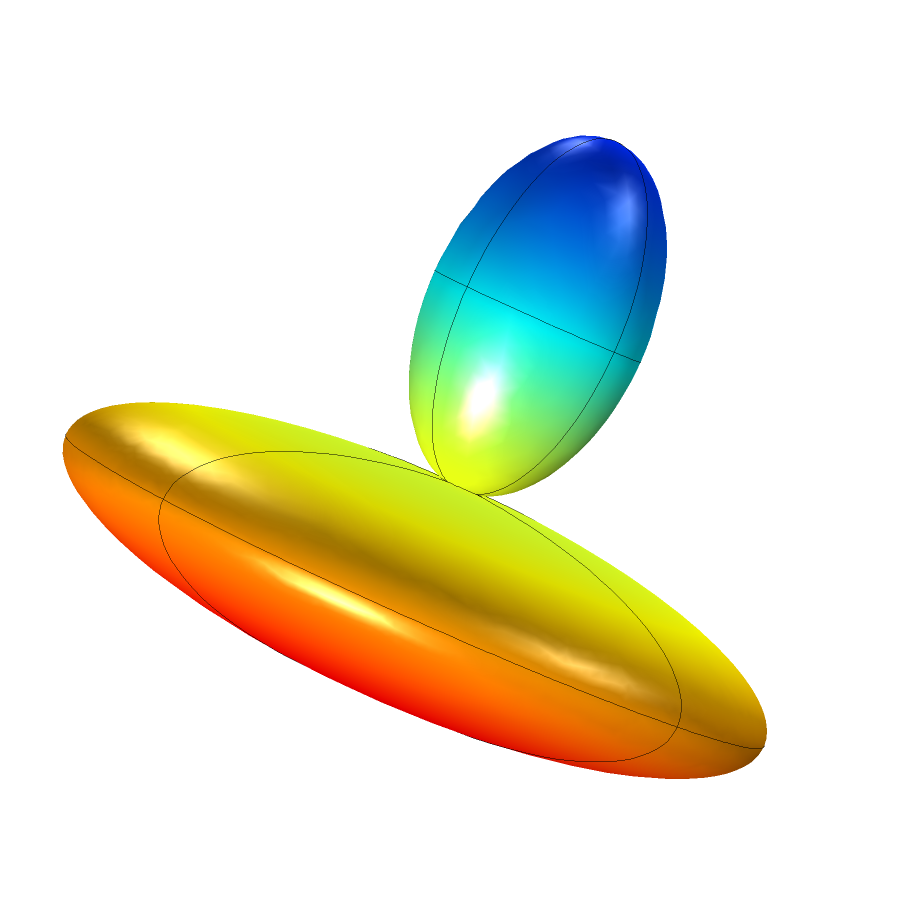}
}
\subfloat[t=10s]
{\label{fig:PolarizationDependingOnShapeEllips3}
 \includegraphics[width=0.2\linewidth]{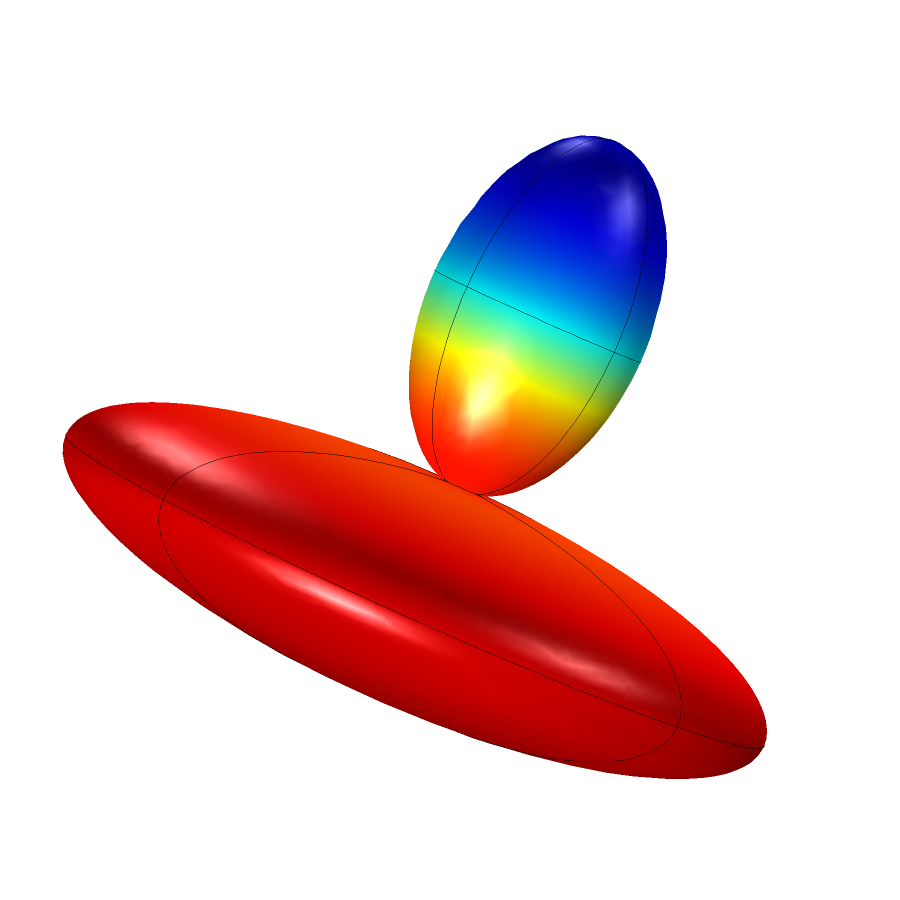}
}
\subfloat[t=100s]
{\label{fig:PolarizationDependingOnShapeEllips4}
 \includegraphics[width=0.2\linewidth]{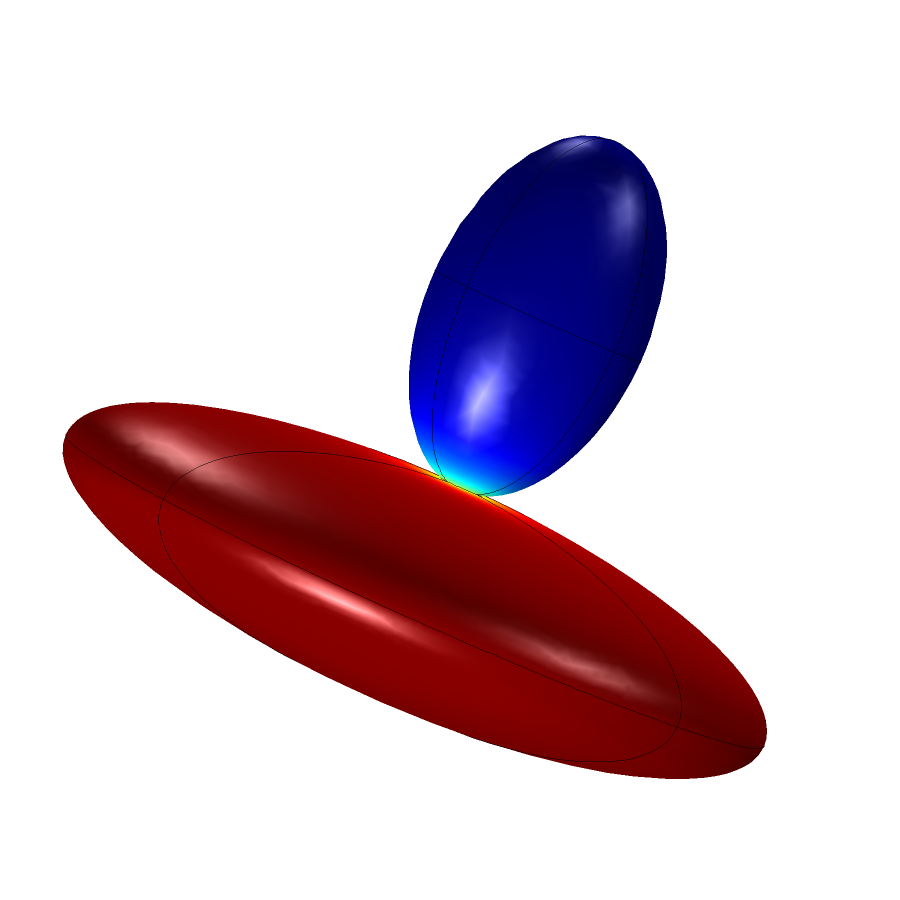}
}
\includegraphics[width=0.13\linewidth]{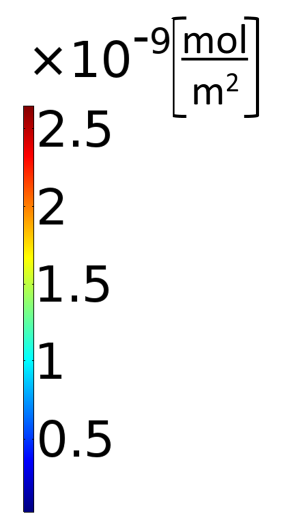}
\caption{\label{fig:PolarizationDependingOnShape}
Active Cdc42 for cells of different shapes: an ellipsoid cell (top
row, (a)-(c)), and a cell composed of two thinly connected ellipsoids (bottom row, (d)-(f)) at times of $5$, $10$, and $100s$.
A stimulus gradient is applied to the initially homogeneous cells. Both cells
initially polarize at $t = 5s$, but the cell with the shape of a single ellipsoid
loses its polarization at $t = 100s$, whereas the cell composed of two ellipsoids
is able to maintain its polarization.
}
\end{figure}

In many experiments, cells present in vastly different shapes. For instance, as
shown in Fig. \ref{fig:RealCells}, a cancer cell is extravasating through a vascular lumen \cite{chen2013mechanisms}. As it does, it transitions from a nearly spherical shape (Fig. \subref*{fig:RealTransmigratingCell1}), into one consisting of a spherical region inside the lumen, spreading into a broad set of protrusions in the extracellular matrix outside of
the lumen (Fig. \subref*{fig:RealTransmigratingCell2}). The two parts of the cell are connected by a thin neck-like region  reaching through the endothelium, barely visible in this single confocal slice, but typically being of about $1$ or $2\mu m$ in diameter. While a full model of the extravsation process would need to take complexities such as the change in environment from the blood-filled inside of the lumen to the extracellular matrix outside of the lumen into account, our current model provides, with limitations, an understanding of what effect complex cell shapes similar to those seen in Figure \ref{fig:RealTransmigratingCell2} would have on the polarization behavior of cells.

We compare a cell with two different shapes: First as a single ellipsoid, Figs.
\subref*{fig:PolarizationDependingOnShapeSphere2}-\subref{fig:PolarizationDependingOnShapeSphere4}, then, as two ellipsoids joined by a thin neck $1.3\mu m$ in diameter between the ellipsoids,
Figs. \subref*{fig:PolarizationDependingOnShapeEllips2}-\subref{fig:PolarizationDependingOnShapeEllips4}. For a better comparison we keep the length and volume of the two configurations the same, so that the main difference between the two cases is
the thinning, and the spreading of one half of the cell, similar as seen in the extravasating cell outside of the lumen in Figure \ref{fig:RealCells}.
We see that for both shapes, the cell is polarizing at $t = 5s$ in response to
the stimulus. However, at $t = 100s$, the single-ellipsoid cell has lost its
polarization, Fig. \subref*{fig:PolarizationDependingOnShapeSphere4}, whereas the extravasating cell maintains a strongly polarized state, Fig. \subref*{fig:PolarizationDependingOnShapeEllips4}, such that active Cdc42 is mainly concentrated in
the part of the cell outside of the lumen. This could explain the formations
of filopodia, known to be directed by Cdc42, almost exclusively outside of
the lumen. However, as mentioned before, the current model does not take all complexities during the extravasation process into account so further work is required to investigate if shape alone, or a combination with other effects such as the presence of ECM molecules outside of the lumen, are responsible for the observed behavior.

%
\subsubsection{Response of Cell to a Change in Stimulus Direction}
\begin{figure}
 \includegraphics[width=0.20\linewidth]{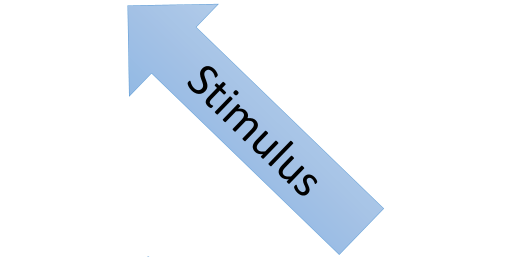}
  \includegraphics[width=0.20\linewidth]{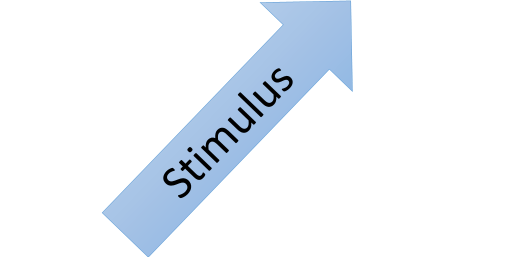}
 \includegraphics[width=0.20\linewidth]{Stimulus_arrow_right_top}
  \includegraphics[width=0.20\linewidth]{Stimulus_arrow_right_top}

\subfloat[t=100s]
{
 \includegraphics[width=0.20\linewidth]{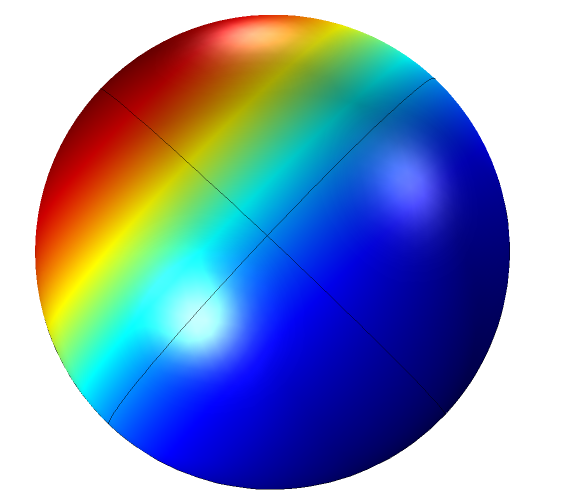}
}
\subfloat[t=140s]
{
 \includegraphics[width=0.20\linewidth]{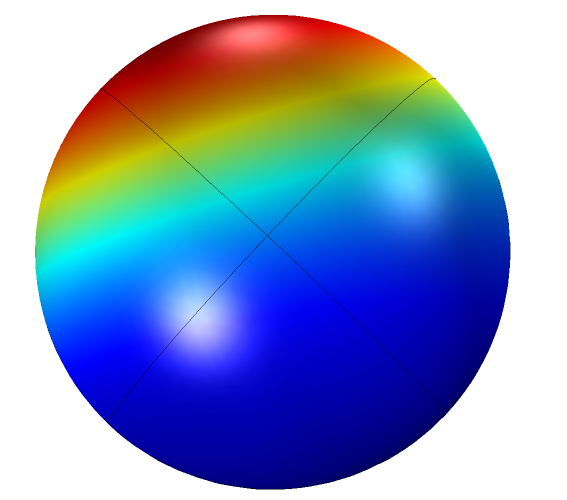}
}
\subfloat[t=180s]
{
 \includegraphics[width=0.20\linewidth]{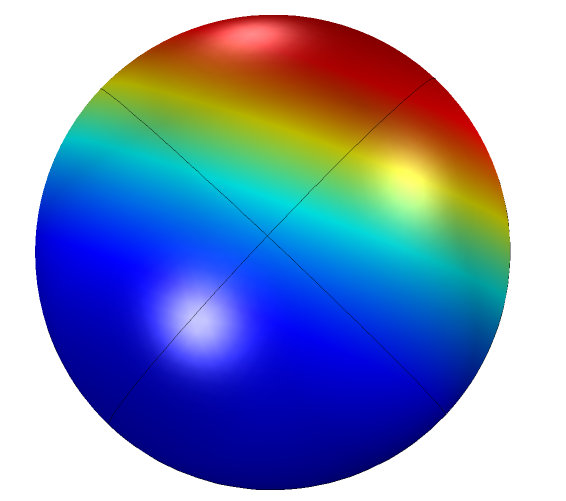}
}
\subfloat[t=300s]
{
 \includegraphics[width=0.20\linewidth]{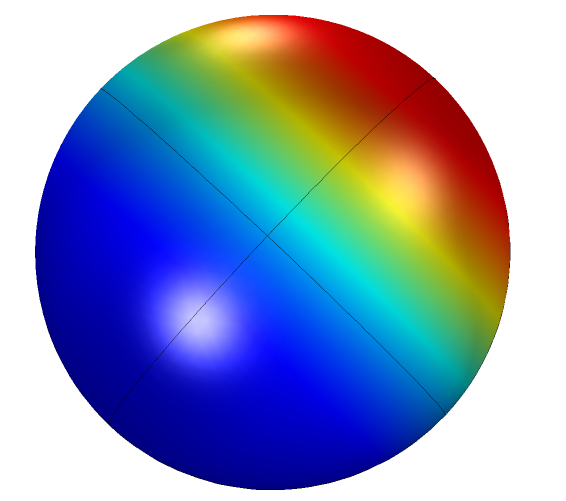}
}
\includegraphics[width=0.13\linewidth]{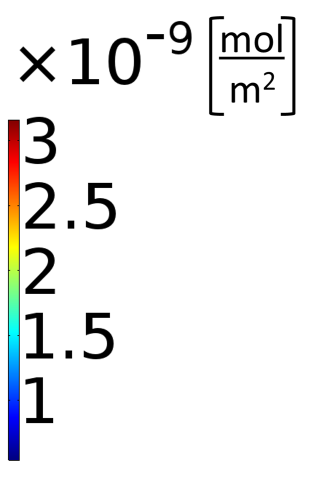}
\\
\subfloat[t=100s]
{
 \includegraphics[width=0.20\linewidth]{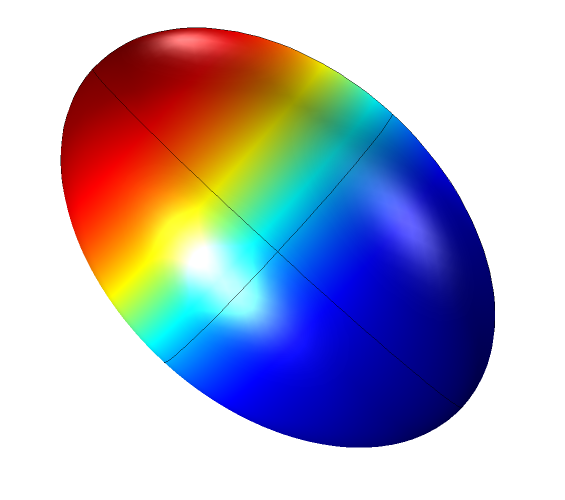}
}
\subfloat[t=140s]
{
 \includegraphics[width=0.20\linewidth]{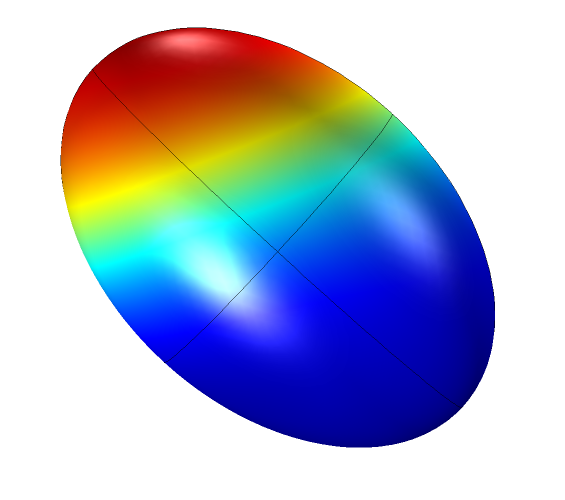}
}
\subfloat[t=180s]
{
 \includegraphics[width=0.20\linewidth]{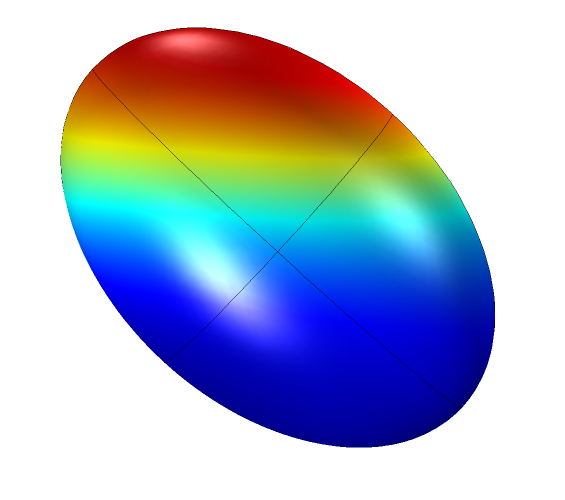}
}
\subfloat[t=300s]
{
 \includegraphics[width=0.20\linewidth]{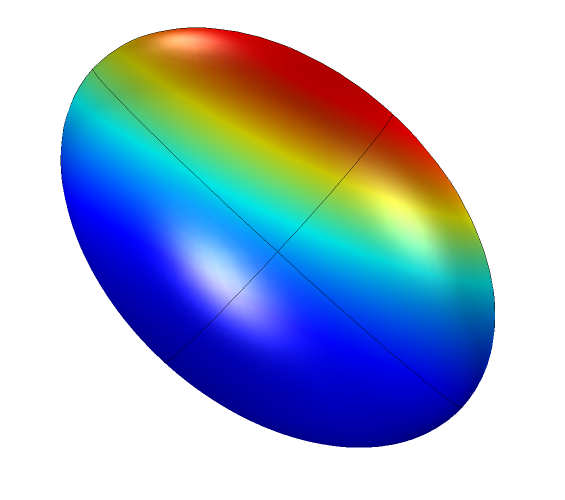}
}
\includegraphics[width=0.13\linewidth]{rotation_legend}
\\
\subfloat[t=100s]
{
 \includegraphics[width=0.20\linewidth]{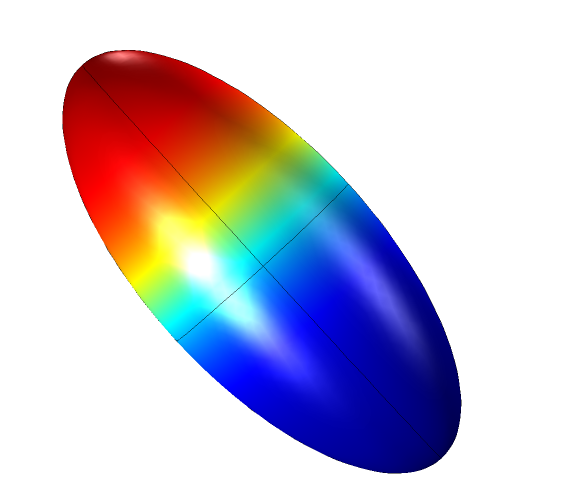}
}
\subfloat[t=140s]
{
 \includegraphics[width=0.20\linewidth]{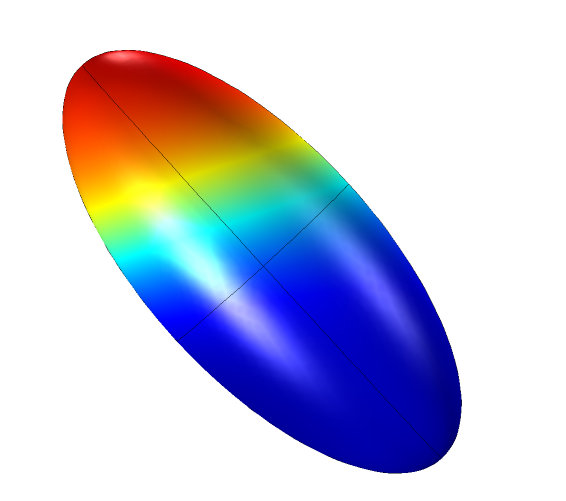}
}
\subfloat[t=180s]
{
 \includegraphics[width=0.20\linewidth]{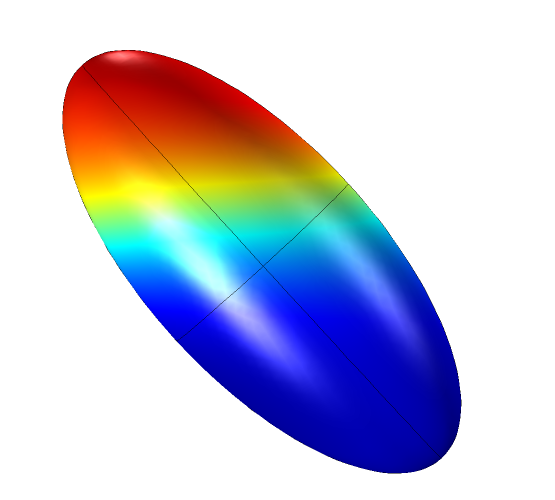}
}
\subfloat[t=300s]
{
 \includegraphics[width=0.20\linewidth]{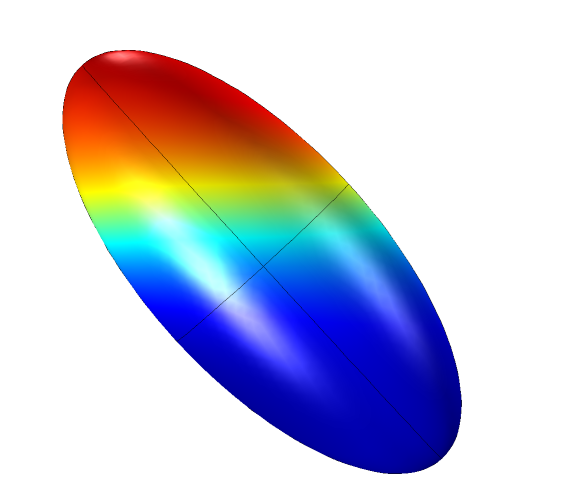}
}
\includegraphics[width=0.13\linewidth]{rotation_legend}
\caption{\label{fig:RepolarizationOfEllipsoids} Active Rac on the membrane is shown at different times for the same cell with different shapes, where the Rac activation rate in the first $100 s$ increases linearly along the long axis of the ellipsoid (from lower right corner to upper left corner), and from then on, it is rotated by $90$ degrees and now increases linearly along a short axis of the ellipsoids (from the lower left corner to the upper right corner). In all cases, the volume of the ellipsoid cells is fixed as $V=800\mu m^3$, the main axis is $11.5\mu m$ (spherical, (a)-(d)), $15\mu m$ ((e)-(h)) and $20\mu m$ ((i)-(l)), and the other two axes are of the same length. Comparing the different shapes, we see that only the spherical cell can completely polarize into the new stimulus direction, whereas the cells with ellipsoidal shapes will form a stable pattern which points into a direction in between the original and final stimulus direction.}
\end{figure}
\begin{figure}

 \includegraphics[width=0.20\linewidth]{Stimulus_arrow_right_top}
 \includegraphics[width=0.20\linewidth]{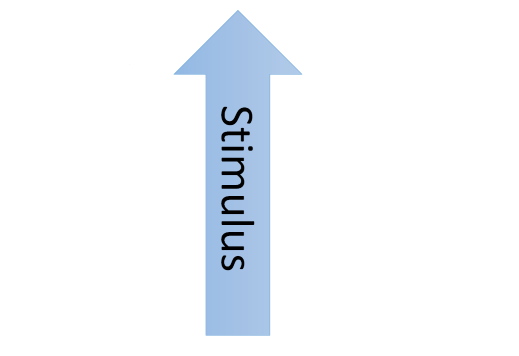}
 \includegraphics[width=0.20\linewidth]{Stimulus_arrow_top_to_bottom}
 \includegraphics[width=0.20\linewidth]{Stimulus_arrow_top_to_bottom}

\subfloat[t=100s]
{
 \includegraphics[width=0.20\linewidth]{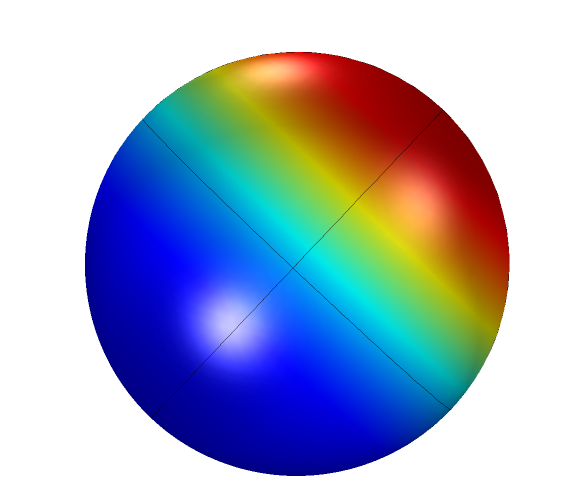}
}
\subfloat[t=140s]
{
 \includegraphics[width=0.20\linewidth]{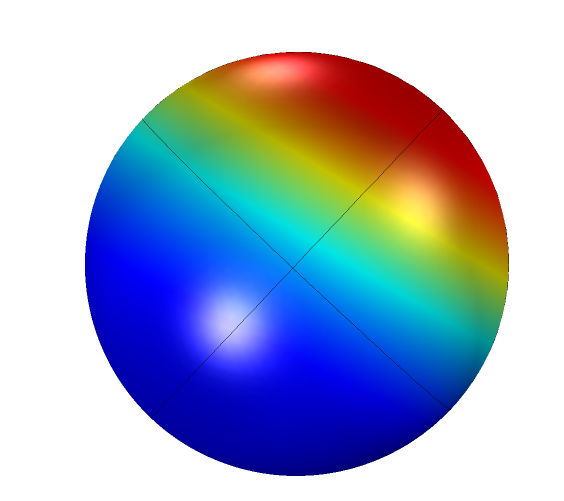}
}
\subfloat[t=180s]
{
 \includegraphics[width=0.20\linewidth]{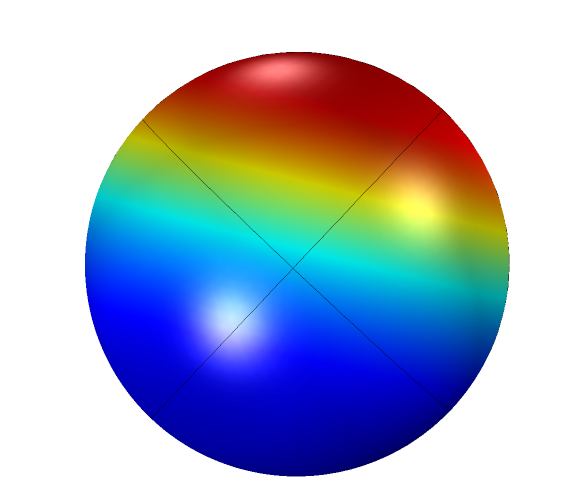}
}
\subfloat[t=300s]
{
 \includegraphics[width=0.20\linewidth]{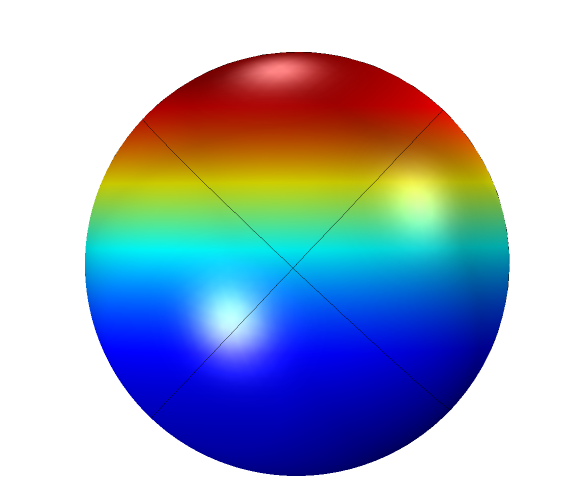}
}
\includegraphics[width=0.13\linewidth]{rotation_legend}
\\
\subfloat[t=100s]
{
 \includegraphics[width=0.20\linewidth]{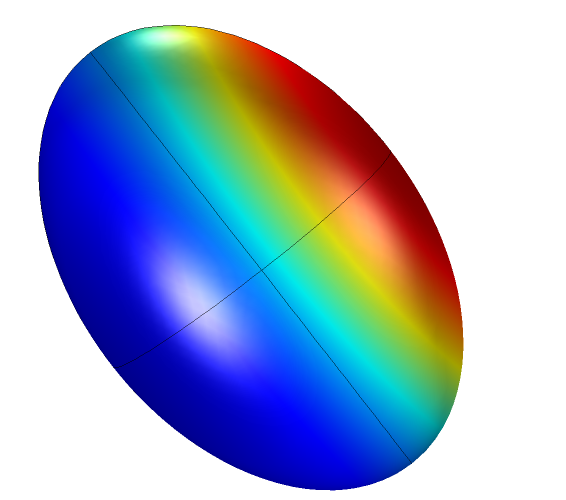}
}
\subfloat[t=140s]
{
 \includegraphics[width=0.20\linewidth]{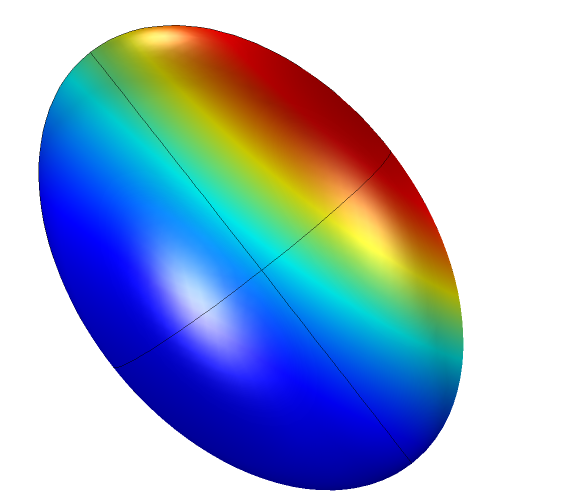}
}
\subfloat[t=180s]
{
 \includegraphics[width=0.20\linewidth]{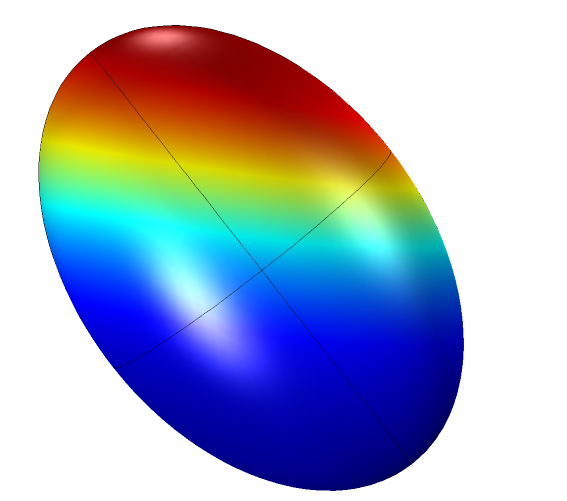}
}
\subfloat[t=300s]
{
 \includegraphics[width=0.20\linewidth]{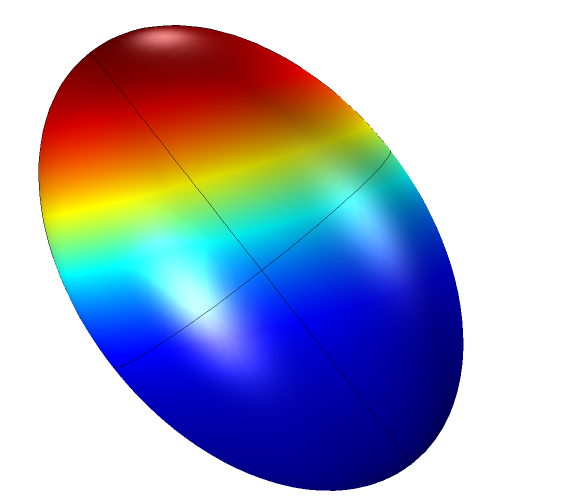}
}
\includegraphics[width=0.13\linewidth]{rotation_legend}
\\
\subfloat[t=100s]
{
 \includegraphics[width=0.20\linewidth]{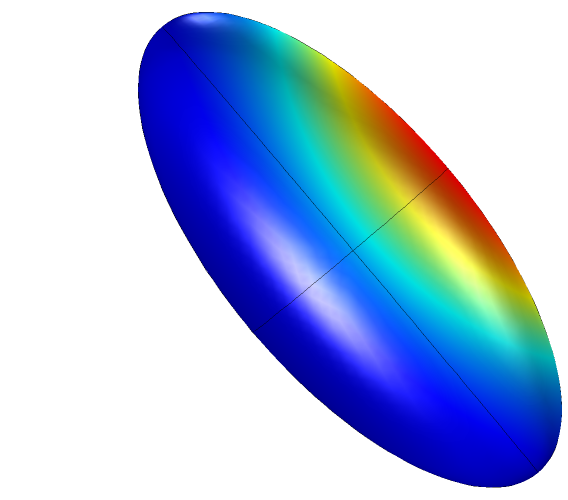}
}
\subfloat[t=140s]
{
 \includegraphics[width=0.20\linewidth]{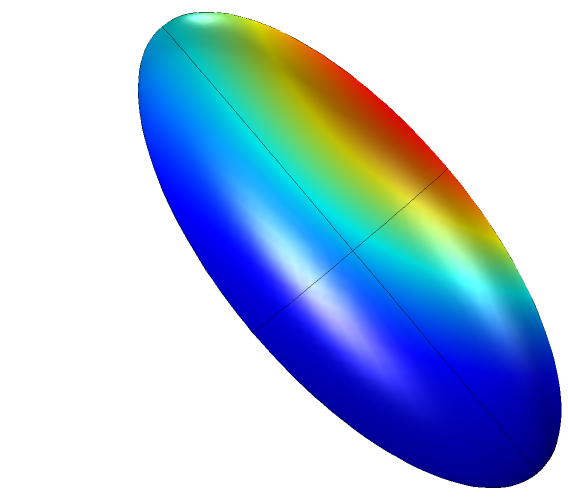}
}
\subfloat[t=180s]
{
 \includegraphics[width=0.20\linewidth]{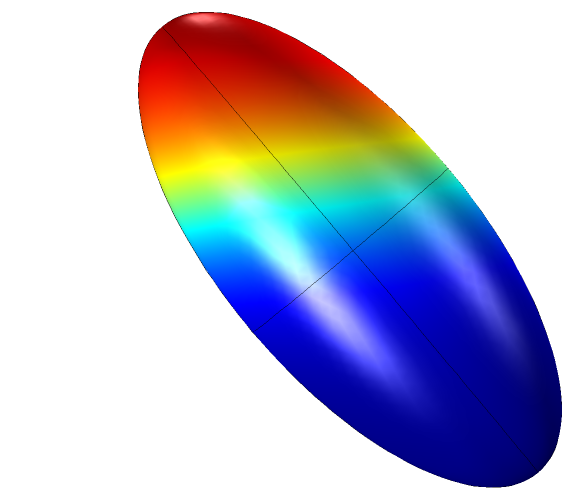}
}
\subfloat[t=300s]
{\label{fig:RepolarizationOfEllipsoidsFromShortAxisEllipsoidFinal}
 \includegraphics[width=0.20\linewidth]{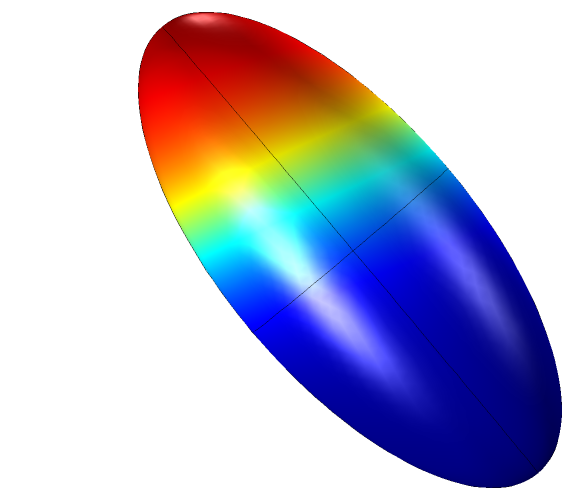}
}
\includegraphics[width=0.13\linewidth]{rotation_legend}
\caption{\label{fig:RepolarizationOfEllipsoidsFromShortAxis}As in Fig. \ref{fig:RepolarizationOfEllipsoids}, active Rac on the membrane is shown for different times and cells of different shapes, but here, the Rac activation rate in the first $100 s$ increases linearly along a short axis of the ellipsoid (from lower left corner to upper right corner), and from then on, it is rotated by $45$ degrees and increases linearly in between the long and a short axis of the ellipsoids (from bottom to top). In all cases, the volume of the ellipsoid cells is fixed as $V=800\mu m^3$, the main axis is $11.5\mu m$ (spherical, (a)-(d)), $15\mu m$ ((e)-(h)) and $20\mu m$ ((i)-(l)), and the other two axes are of the same length. Comparing the different shapes, we see that the spherical cell can completely polarize into the new stimulus direction, whereas the cells with the elongated ellipsoidal shapes will rotate their internal polarization direction further than the stimulus direction, so that the polarization is more aligned with their long axis.}
\end{figure}
We now investigate how cell shape can influence the response of a cell to a change in the direction of a stimulus. This will allow us to make predictions with this model beyond the response to unidirectional stimuli, which exist typically in {\it in vitro} setups such as classical Boyden chambers \cite{seppa1982platelet}, or modern microfluidic platforms \cite{shamloo2008endothelial,zervantonakis2010concentration,chung2010microfluidic,kim2013cooperative}. However, many cases of cell migration {\it in vivo} are more complex due to tissue heterogeneity as well as temporal and spatial changes in stimuli \cite{cai2014diverse}. The scenario presented in Fig. \ref{fig:RepolarizationOfEllipsoids} goes towards an understanding of how cells respond to changes in stimuli, and how this response is affected by cell shape. Active Rac is shown for a cell of a volume of $V=800\mu m^3$ for three different ellipsoidal configurations: With a main axis of $11.5\mu m$ (spherical, (a)-(d)), $15\mu m$ ((e)-(h)) and $20\mu m$ ((i)-(l)).
Initially a stable polarized state is obtained in all three configurations, which is shown at $t=100s$ just before the activation gradient direction is changed. The spherical cell always maintains a main direction, but the direction of polarization rotates towards the new stimulus direction. At $t=180 s$ the cell is mainly polarized into the new stimulus direction (c), and at $t=300 s$ it is completely repolarized. Interestingly, the ellipsoidal configurations never reach a polarized state aligned with the new stimulus. Instead, the new polarization direction is somewhere in between the original and final stimulus direction, and the larger the ratio of the ellipsoid axes is, the closer the new polarization direction will remain to the original polarization direction. This would suggest that a more symmetric, spherical cell is able to repolarize faster and more efficiently than a long, thin cell, despite their chemical pathways being unaltered. In Fig. \ref{fig:RepolarizationOfEllipsoidsFromShortAxis}, we use the same setup as in Fig. \ref{fig:RepolarizationOfEllipsoids}, but we initially polarize along one of the short ellipsoid axes and then rotate by $45$ degrees towards the long axis. The spherical cell can, as before, adapt to the new stimulus direction. Now, the ellipsoidal cells can rotate faster, and, interestingly, they rotate their internal polarization state by more than $45$ degrees towards the long axis. Indeed, the longest ellipsoidal cell is, after $t=300s$, almost fully polarized along the long axis (Fig. \subref*{fig:RepolarizationOfEllipsoidsFromShortAxisEllipsoidFinal}). Supplementary Fig. \ref{fig:RepolarizationOfEllipsoidsFromShortAxisFullRotation} shows the same setup, where, like in Fig. \ref{fig:RepolarizationOfEllipsoidsFromShortAxis} we initially polarize along a short axis, but then fully rotate by $90$ degrees towards the long axis. Here, all cells can repolarize towards the new direction, and the ellipsoidal cells repolarize faster than the spheroidal cells.
It is well known in the literature that roundish, amoeboid cells can quickly polarize and adapt to new stimuli, in contrast to more elongated cells such as mesenchymal cells \cite{vorotnikov2014chemotactic}. Futhermore, in 3D matrices, elongated mesenchymal cells can migrate persistently along the fibrous structures. Whereas traditionally mesenchymal and amoeboid cells denoted different cell types, such as fibroblasts on the one hand, and dictyostelium or neutrophils on the other hand, recent work has also focused on a switch between those migratory modes for the same cell types. Most of this work has focused on alterations of biochemical pathways to describe the switch between mesenchymal and amoeboid migrations. In \cite{wolf2003compensation}, the role of MMPs in this switch was investigated, in \cite{sahai2003differing} an important role  of Rho and ROCK was discovered, and in \cite{mishima2010lim} LIM kinase was implicated in this switch. In \cite{huang2015interstitial}, cell aspect ratios were taken as the factor determining whether a cell migrates in a mesenchymal or amoeboid way, with the mesenchymal cell being more persistent, and interstitial flow was shown to affects the switch between these migratory modes. Here, our argument shows that, without any changes in the biochemical pathways, roundish cells are expected to repolarize towards a new stimulus more quickly than elongated cells purely because of their different shape. On the other hand, elongated cells preferentially migrate in the direction of their longest extent, even if the stimulus appears in a slightly different direction, giving them an increased persistence.

\section{Conclusion}
Motivated by the fact that cells, both in {\it in vitro} and {\it in vivo} environments, present with greatly varying shapes, in this paper we have investigated how cell shape influences gradient detection and cell polarization. For molecules such as small GTPases, which can exist both in a membrane bound form as well as sequestered in the cytosol, the ratio of cell volume to surface area can influence important properties such as the fraction of membrane-bound molecules, as seen in equation \eqref{eq:fractionBoundGeneral}. Similar arguments were provided in \cite{meyers2006potential}. However, all GTPases are affected in this way, and they are partially inhibitory to each other, so it is not a priori clear how a full model with interacting molecules would behave under change of cell shape. Furthermore, we found that even in the dimensionally reduced models, the higher dimensions implicitly appeared in the reduced equation by renormalizing the membrane binding rate (equation \eqref{eq:cylinderReductionTo1D}). We also found conditions on the parameters which, when satisfied, justify the use of the lower dimensional models (equation \eqref{eq:condition1Dmodel1}). These conditions are quite general and should be useful both to check the validity of older models as well as for the development of new models which involve the interactions of molecules between the membrane and the cytosol.

In our model, we have only considered constant binding and unbinding rates of a molecule to and from the membrane, and investigated how these rates affect the polarization behavior of a cell (Fig. \ref{fig:offRateDependence}). In real cells, the binding and unbinding of small GTPases to and from the membrane as well as their activation and deactivation are influenced by the presence of a large number of different molecules such as GDIs, guanine nucleotide exchange
factors (GEFs) and GTPase activating proteins (GAPs) \cite{garcia2011invisible}. It would be interesting to include the effect of these regulating molecules in our model. However, at present, there is a lack of good quantitative data regarding the spatio-temporal regulation of these molecules, so we postpone such investigations for future research. What Fig. \ref{fig:offRateDependence} confirmed is that the binding/unbinding rate can influence the ability of a cell to polarize, and as GDI molecules are expected to modify those rates the model predicts that the presence or absence of these molecules will also affect the polarization behavior.

We then extended an established cell polarization pathway \cite{holmes2012modelling}, which was previously investigated in a 1D model of HeLa cells, to our 3D model. The purpose of choosing \cite{holmes2012modelling} for comparison was that in this model, the effect of changing the cell length was implicitly taken into account via a modification of the fraction of membrane-bound GTPases, whereas most other 1D models did not consider any geometric effects at all. Hence, a first test of our model was to reproduce some results of \cite{holmes2012modelling} and highlight quantitative and qualitative differences. Furthermore, as mentioned above, we are, to our knowledge, the first to explicitly consider the binding/unbinding dynamics of GTPases to the membrane and show the influence of those parameters on the polarization behavior of the cell in Fig. \ref{fig:offRateDependence}. 

We then explored scenarios which the existing lower dimensional models could not capture. First, we compared the polarization behavior of two cells with the same volume and length, one ellipsoid, and one cell composed of two connected ellipsoids, Fig. \ref{fig:PolarizationDependingOnShape}. The second shape was motivated by shapes observed during cancer cell extravasation \cite{chen2013mechanisms}, where the cellular environment can impose different extreme shapes on the cells. This result is important whenever one is trying to compare theoretical results obtained from simplified lower dimensional models to experiments, indicating that one has to take cell shape into account. We expect that the behavior of pathways other than those describing polarization would also be affected by cell shape in a similar manner, if the principle mechanism of polarization is mediated by similar reaction-diffusion models as the ones used here.

As long as the polarization stimulus is coming from only one direction, and provided the parameter constraints \eqref{eq:condition1Dmodel1} are satisfied, 1D models could still be derived which take into account if the cross section of the cell along the stimulus direction is relatively constant. However, purely one-directional stimuli are idealistic and {\it in vivo} different stimuli can appear from all directions and change in time. In the study of such effects we have to use 3D models such as ours. In Fig. \ref{fig:RepolarizationOfEllipsoids} we simulated a stimulus which was changing directions over time. We found that cell shape is an important factor which predicts how fast and strong a cell can adapt to the new direction of stimulus. Indeed, the spherical cell was able to change the internal polarization direction smoothly towards the new stimulus direction, whereas cells which are elongated along the previous direction of stimulus preferentially stayed polarized in a direction close to the original stimulus. This is compatible with experimental findings that roundish amoeboid cells are typically faster to adapt to new stimuli than mesenchymal cells, which are typically more elongated. An explicit test of the model prediction could be conducted, for instance, using a microfluidics platform where one can change the direction of an external growth factor gradient over time, and measure the response of some tagged internal molecule associated with polarization for varying cell shapes. This would be a step toward an understanding of cell polarization under temporally and spatially varying conditions as typically present {\it in vivo} \cite{cai2014diverse}. From a theoretical point of view, it would be interesting to include the effect of dynamical changes of shape through coupling of mechanics with our biochemical pathways, as these dynamical changes have also been shown to affect polarization behavior on longer time-scales in a 2D model \cite{maree2012cells}.

In summary, the results in this paper predict the importance of cell shape on polarization of cells, indicate in which cases the use of lower dimensional models is justified, and demonstrate when a full 3D model such as ours needs to be used to model and predict experimental results.
%

\section*{Acknowledgment}
We acknowledge the support of the NCI grant number 5U01CA177799. We are grateful to Leah Edelstein-Keshet, Eamonn Gaffney, Bill Holmes, Philip Maini, Noppadol Mekareeya, Fernando Santos, Robert Seager and Frits Veerman for useful discussions, and the reviewers for helpful suggestions.

%
%

\newpage
\beginsupplement

\section*{Supplementary Information}

\section{Derivation and Consistency of the Membrane Binding-Unbinding Model}\label{sec:App:MembraneCytosolBinding}

We model a molecule which is diffusing in a cell, can bind to and unbind from the cell membrane, and diffuse on the membrane when bound. Let $\Gc$ be the concentration of this molecule in the cytosol, i.e. the inside of a cell. We denote the domain of the cytosol by $V\subset\mathbb{R}^3$, which is a smooth Riemannian manifold with the metric induced from the Euclidean metric in $\mathbb{R}^3$. Likewise, $\Gm$ is the concentration of the same molecule when bound to the membrane, which is defined as the boundary of $V$, $S=\partial V$, and is an orientable Riemannian manifold. The membrane-bound molecules can unbind, and the molecules in the cytosol can bind to the membrane, with rates $k_{on}$ and $k_{off}$, and where $L_I$ is a length scale associated with the binding range of a sequestered molecule to the membrane. Furthermore, $D_M$ and $D_C$ denote the diffusion coefficients for diffusion on the membrane and in the cytosol, respectively. Our equations are given by \eqref{eq:membraneCytosolBasicEquation}, which we repeat here for convenience:
\begin{align}\label{eq:App:membraneCytosolBasicEquation}
\frac{\partial \Gm({\bar r_m},t)}{\partial t} &= D_M \nabla_S^2 \Gm({\bar r_m},t) + k_{on}L_I \Gc({\bar r_m},t) - k_{off}\Gm({\bar r_m},t),\nonumber\\
\frac{\partial \Gc({\bar r_c},t)}{\partial t} &= D_C \nabla_V^2\Gc({\bar r_c},t),\nonumber\\
-D_C e_n \nabla_V \Gc({\bar r_m},t) &= k_{on}L_I \Gc({\bar r_m},t) - k_{off}\Gm({\bar r_m},t), \nonumber\\
\Gm({\bar r_m},0) &= \Gm^0({\bar r_m}), \nonumber\\
\Gc({\bar r_c},0) &= \Gc^0({\bar r_c}).
\end{align}
Here, $\nabla_S^2, \nabla_V^2$ denote the Laplace operators (otherwise denoted as Laplacian, or Laplace-Beltrami operator) on $S$ and $V$, respectively, and are defined in the usual way on Riemannian manifolds \cite{jost2008riemannian}. Furthermore, $e_n$ denotes the uniquely defined unit outwards normal vector on the surface, and ${\bar r_c}\in V$, ${\bar r_m}\in S$. Hence, $e_n \nabla_V \Gc({\bar r_m},t)$ denotes the projection of the gradient of $\Gc$ on the unit normal vector on the surface. We have imposed the outwards normal flux in such a way that it matches the binding and unbinding reactions and preserves total particle numbers. Furthermore, $ \Gm^0({\bar r_m}), \Gc^0({\bar r_c})$ denote functions defining the initial conditions, and naturally $\Gm$ does not need any boundary conditions, as it is defined on a surface without boundary.

\subsection{Particle Number Conservation}
We now show that the equations given in \eqref{eq:App:membraneCytosolBasicEquation} conserve the number of particles. The total amount of molecules is given by 
\begin{align}
N &= \int_V \Gc\, dV + \int_S \Gm\, dS.
\end{align}
This total amount of molecules is conserved by choice of boundary condition:
\begin{align}
\frac{\partial N}{\partial t} &= \int \frac{\partial \Gc}{\partial t} dV + \int \frac{\partial \Gm}{\partial t} dS\nonumber\\
&= \int D_C div_V (grad_V \Gc) dV + \int (D_M div_S (grad_S \Gm) + k_{on}L_I \Gc - k_{off}\Gm)dS\nonumber\\
&= \int D_C e_n (grad_V \Gc) dS + \int (D_M div_S (grad_S \Gm) + k_{on}L_I \Gc - k_{off}\Gm)dS\nonumber\\
&= \int D_C e_n (grad_V \Gc) dS + \int (0 + k_{on}L_I \Gc - k_{off}\Gm)dS\nonumber\\
&= \int(-k_{on}L_I \Gc + k_{off}\Gm+ k_{on}L_I \Gc - k_{off}\Gm)dS\nonumber\\
&= 0
\end{align}
In the first line of the derivation, we have simply plugged in the time derivatives of $\Gc$ and $\Gm$ from equation \eqref{eq:membraneCytosolBasicEquation}. Here, we have used that (on Riemannian manifold, independent of the coordinate system) we can write the Laplace operator as divergence of a gradient, where the subindices indicate the corresponding manifold in which divergence or gradient are calculated. We do not need the precise definition of gradient, divergence or Laplacian on those manifold, we only need the fact that the divergence theorem applies. Indeed, we apply the divergence theorem in the second step, changing from an integral of a divergence of the gradient of $\Gc$ over the whole cell to an integral of the normal flux over the boundary. Then, in the third step, we apply the divergence theorem to the divergence of the gradient of $\Gm$. However, the surface does not have a boundary, so the divergence theorem immediately gives zero for this term. In the final step, we plug in the boundary condition from equation \eqref{eq:membraneCytosolBasicEquation} for the normal flux of $\Gc$ at the boundary, and obtain our final result, that the total amount of particles is conserved.

\subsection{Global Invariance of Boundary Condition}
Global conservation of particles is still ensured by adding a Laplacian of $\Gm$ to the flux boundary conditions:
\begin{align}\label{eq:alternative}
 -D_C e_n \nabla_V \Gc({\bar r_m},t) &= k_{on}L_I \Gc({\bar r_m},t) - k_{off}\Gm({\bar r_m},t) + \lambda\nabla_S^2 \Gm({\bar r_m},t)
\end{align}
It is immediately clear that this preserves total particle numbers for any $\lambda$, for if we integrate the boundary condition over the whole boundary, this term drops out by application of the divergence theorem over a manifold with empty boundary. However, for local conservation, we should not keep this term: If we choose $k_{on}=k_{off}=0$, then $m$ does not bind or unbind from the membrane at all and is conserved on its own (and not just the sum of bound and unbound molecules). However, keeping boundary condition \eqref{eq:alternative} with $\lambda\neq 0$ would result in a flux of $c$ even in that case. On physical grounds, we have to impose Neumann no-flux boundary conditions and the right-hand side of \eqref{eq:alternative} should be zero. Hence, only $\lambda=0$ ensures local conservation of particles.

\subsection{Derivation from a Model with Finite Binding Radius}

We can consider a generalization of equations \eqref{eq:membraneCytosolBasicEquation}, \eqref{eq:App:membraneCytosolBasicEquation} where the change of the membrane density $m({\bar r_m},t)$ is affected by all molecules in the cytosol within a finite radius $L_I$ from the point ${\bar r_m}$. Then, the membrane binding-unbinding model is described by the equations
\begin{align}\label{eq:App:membraneCytosolGeneralEquation}
 \frac{\partial \Gm({\bar r_m},t)}{\partial t} &= D_M \nabla_S^2 \Gm({\bar r_m},t) \nonumber\\
 &+ {\tilde k_{on}}\int_{|{\bar r_m}-{\bar r_c}|\leq L_I} \Gc({\bar r_c},t) - k_{off}\Gm({\bar r_m},t)\nonumber\\
\frac{\partial \Gc({\bar r_c},t)}{\partial t} &= D_C \nabla_V^2\Gc({\bar r_c},t), \nonumber\\
 -D_C e_n \nabla_V \Gc({\bar r_m},t) &=  {\tilde k_{on}}\int_{|{\bar r_m}-{\bar r_c}|\leq L_I} \Gc({\bar r_c},t) - k_{off}\Gm({\bar r_m},t).
\end{align}
We have assumed that all molecules which are with a distance of $L_I$ to a point on the membrane ${\bar r_m}$ are equally likely to be bound with a rate of ${\tilde k_{on}}$, which could be generalized further by including a kernel in the integral such that molecules closer to the membrane are more likely to bind. However, if we assume that $L_I$ is small (see the discussion on parameters in the supplementary information \ref{sec:App:parameters}) so that $\Gc$ does not significantly vary on this length scale, and that the membrane is not significantly curved on this scale, we can reduce the integral terms in \eqref{eq:App:membraneCytosolGeneralEquation}. The integral will thus be over a half-sphere with radius $L_I$, so we can simplify ${\tilde k_{on}}\int_{|{\bar r_m}-{\bar r_c}|\leq L_I} \Gc({\bar r_c},t)= {\tilde k_{on}}\frac{1}{2}\frac{4}{3}\pi L_I^3\Gc({\bar r_m},t)= k_{on}L_I\Gc({\bar r_m},t)$, where we have identified $k_{on} = \frac{2}{3}\pi L_I^2{\tilde k_{on}}$. Hence, \eqref{eq:App:membraneCytosolGeneralEquation} reduces to equations \eqref{eq:membraneCytosolBasicEquation}, \eqref{eq:App:membraneCytosolBasicEquation}.

\subsection{Alternative Derivation from a Discrete Model}

\begin{figure}[htp]
\begin{center}
\begin{tabular}{cc}
\includegraphics[width=0.48\textwidth]{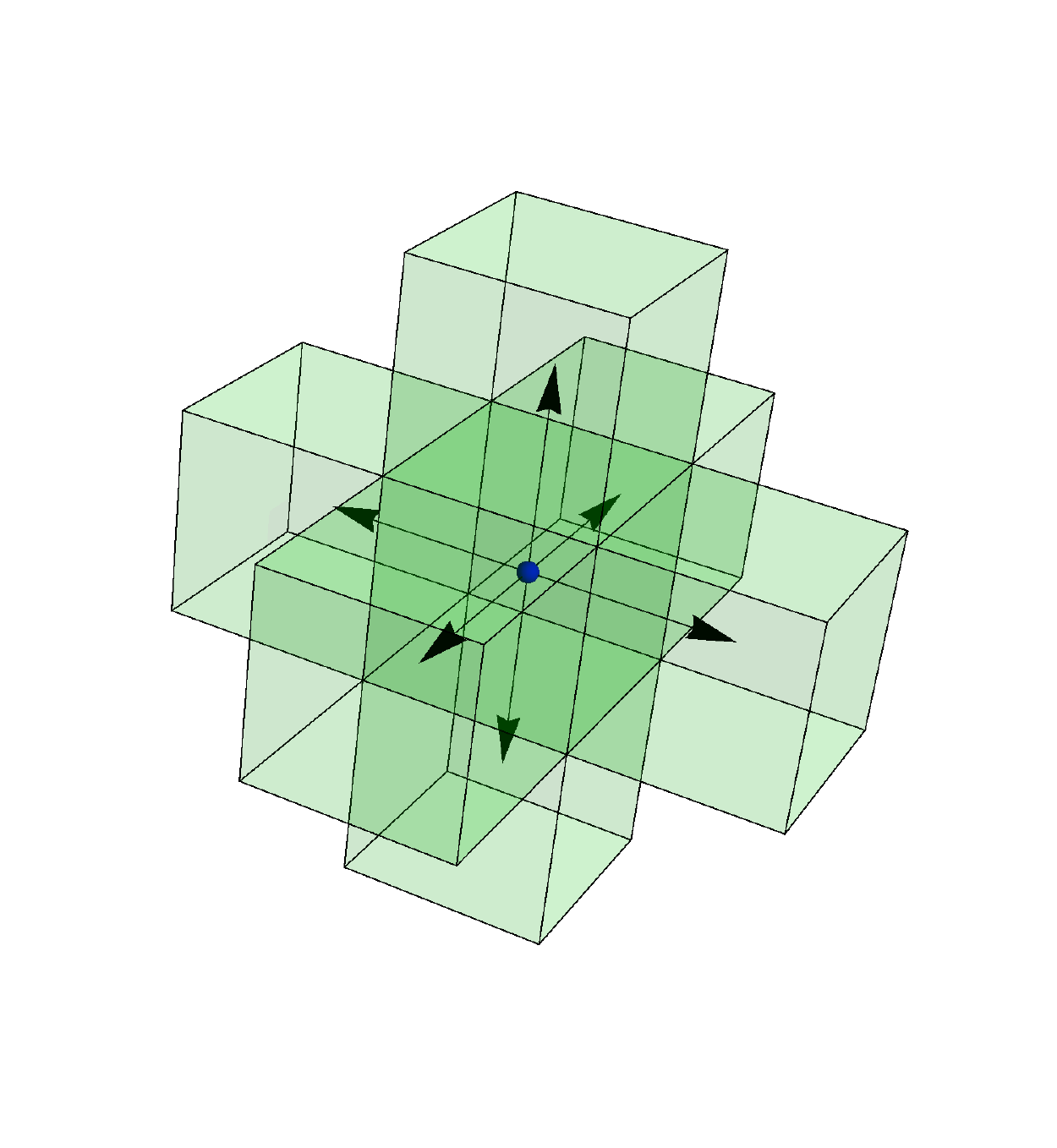} &
\includegraphics[width=0.48\textwidth]{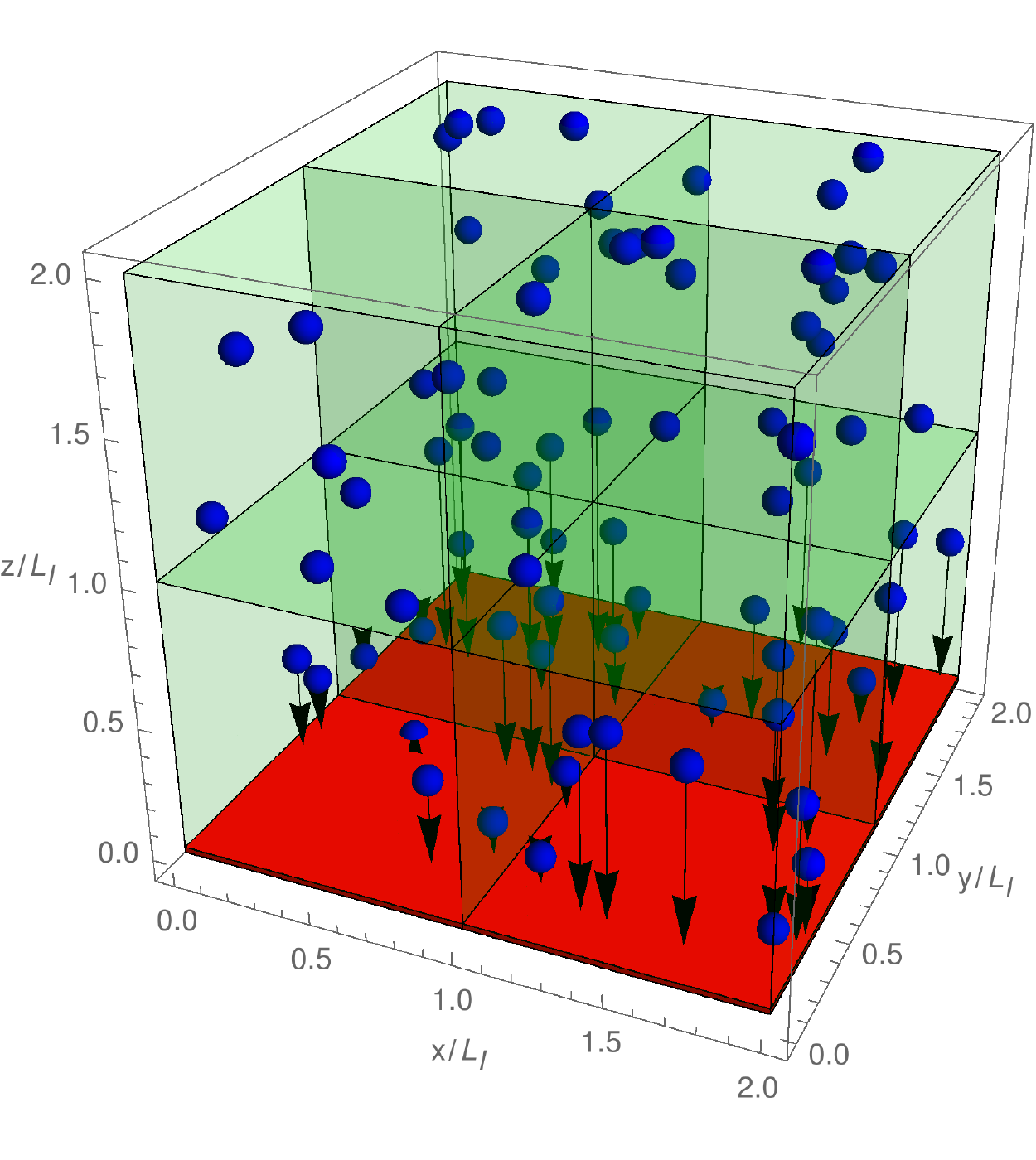} \\
(a) Diffusion in cytosol & (b) Binding to the membrane\\
\includegraphics[width=0.48\textwidth]{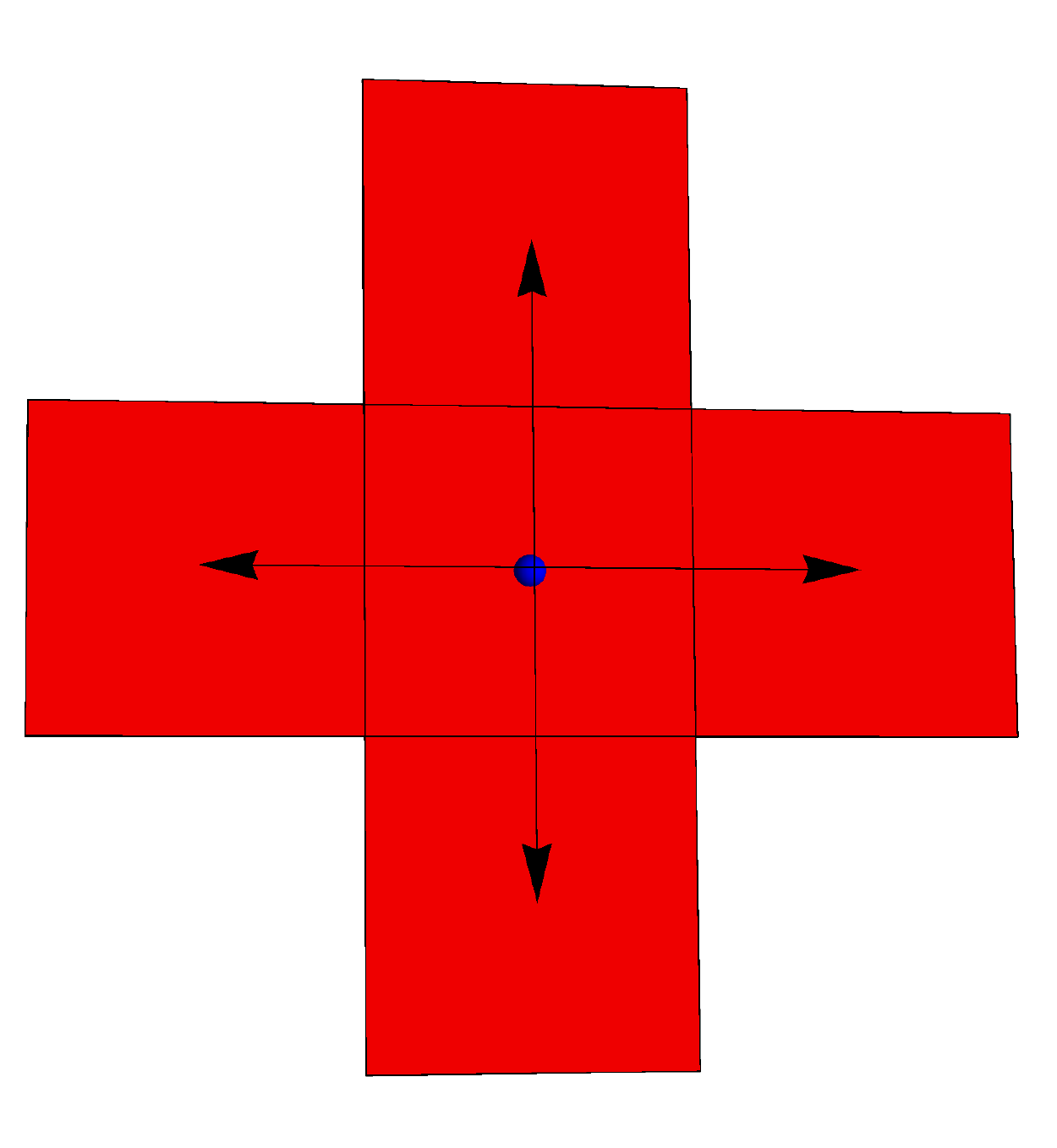} &
\includegraphics[width=0.48\textwidth]{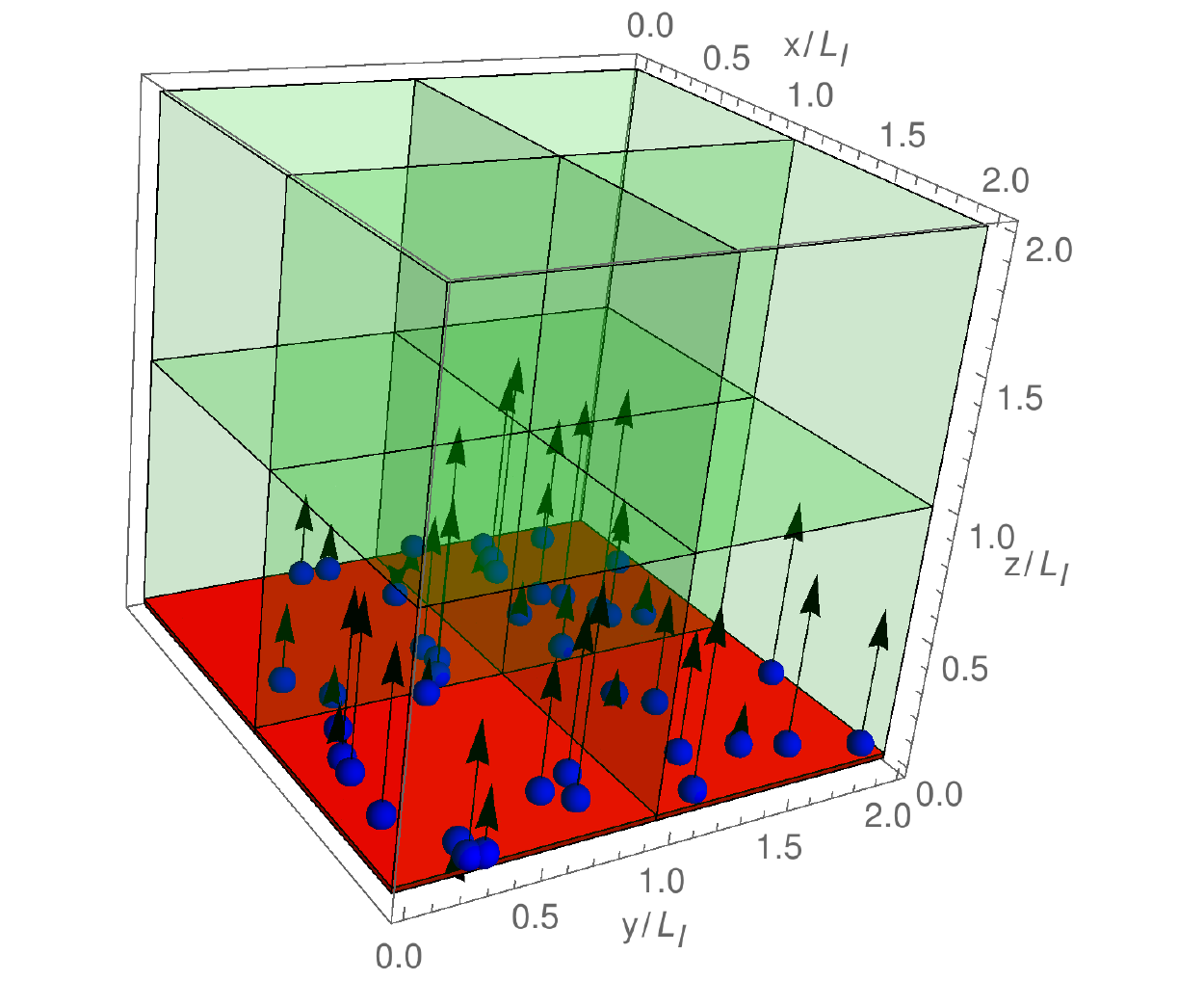} \\
(c) Diffusion on membrane & (d) Unbinding from the membrane
\end{tabular}
\caption{
Molecules in the cytosol can move from their current cube to any of the nearest-neighbor cubes (figure (a)), or, if they are within a distance of the interaction range $L_I$ to the membrane, they can bind to the membrane (figure (b)). Membrane-bound molecules can diffuse on the membrane only (figure(c)), or unbind from the membrane (figure (d)).
}
\label{fig:DiffusionAndMembraneBinding}
\end{center}
\end{figure}

We now give an alternative derivation of equation \eqref{eq:membraneCytosolBasicEquation} from a discrete model. We consider a small section of a cell near the cell membrane, so small that we can ignore the curvature of the membrane. Such section is shown in a schematic drawing on the right panel of Fig.~\ref{fig:DiffusionAndMembraneBinding}, where the cell membrane is highlighted by the red surface. We are interested in the dynamics of the binding and unbinding of molecules to the membrane. Let $L_I$ be the interaction length such that when a molecule in the cytosol is within a distance less or equal to $L_I$ of the membrane, there is a probability of binding this molecule to the membrane. The associated binding rate is denoted by $k_{on}$. Likewise, unbinding is denoted $k_{off}$.

We define the domain of interest to be a cube of length $L\gg L_I$, which we discretize into equally spaced small cubes of size $\delta$. Initially, we identify $\delta=L_I$. Each cube is labeled by integer-valued indices $(k,l,p)$, and the membrane is located at the boundary $p=0$. Then $M(k,l)$ denotes the number of membrane-bound molecules at the membrane segment adjacent to cube $(k,l,0)$, and $C(k,l,p)$ denotes the number of cytosolic molecules in the cube $(k,l,p)$. We consider the following processes: In the inner part of the cytosol, unbound molecules can diffuse only. At the cube adjacent to the membrane, they can diffuse in parallel to the membrane or away from the membrane, or they can bind to the membrane. On the other hand, membrane-bound molecules can unbind, or diffuse on the membrane.
The following equations describe the rate of changes of the average number of molecules:
\begin{align}
\label{eq:meanFieldEquations}
 \frac{\partial M(k,l)}{\partial t} &= \frac{D_M}{\delta^2}\left(M(k+1,l)+M(k-1,l)+M(k,l+1)+M(k,l-1)-4M(k,l)\right) + \nonumber\\
&\quad k_{on}C(k,l,0)-k_{off}M(k,l),\nonumber\\
\frac{\partial C(k,l,p)}{\partial t} &= \frac{D_C}{\delta^2}(C(k+1,l,p)+C(k-1,l,p)+C(k,l+1,p)+C(k,l-1,p)\nonumber\\
&\quad +C(k,l,p+1)+C(k,l,p-1)-6C(k,l)),\quad p>0,\nonumber\\
\frac{\partial C(k,l,0)}{\partial t} &= \frac{D_C}{\delta^2}(C(k+1,l,0)+C(k-1,l,0)+C(k,l+1,0)+C(k,l-1,0)\nonumber\\
&+C(k,l,1)-5C(k,l)) + k_{off}M(k,l)-k_{on}C(k,l,0)\,.
\end{align}
We now want to study the continuum limit of those equations. First, we add an artificial layer of cubes at $p=-1$, such that 
\begin{equation}
\label{eq:meanFieldBoundaryCondition}
C(k,l,-1) := \frac{\delta^2}{D_c}\left(k_{off}M(k,l)-k_{on}C(k,l,0)\right)+C(k,l,0)\,.
\end{equation}
The benefit of this layer is that we can now combine the last two equations of \eqref{eq:meanFieldEquations} into
\begin{align}
 \frac{\partial C(k,l,p)}{\partial t} &= \frac{D_C}{\delta^2}(C(k+1,l,p)+C(k-1,l,p)+C(k,l+1,p)+C(k,l-1,p)\nonumber\\
&\quad +C(k,l,p+1)+C(k,l,p-1)-6C(k,l)),\quad p>-1\,,
\end{align}
which is supplemented by \eqref{eq:meanFieldBoundaryCondition}. From those equations, it is straight-forward to take the continuum limit $\delta\to 0$. We define $\Gc(x,y,z)=\frac{C(k,l,p)}{\delta^3}$ to be the density of unbound molecules at a point $(x,y,z)$, where we identify $(x,y,z)=(\delta k,\delta l,\delta p)$, and likewise $\Gm(x,y)=\frac{M(k,l)}{\delta^2}$ is the surface density of molecules bound to the membrane. Then, the continuum limit $\delta\to 0$ gives the following equation:
\begin{align}
 \frac{\partial \Gm(x,y)}{\partial t} &= D_M \nabla_S^2 \Gm(x,y) + k_{on}\int_0^{L_I}\Gc(x,y,z)dz - k_{off}\Gm(x,y)\nonumber\\
\frac{\partial \Gc(x,y,z)}{\partial t} &= D_C \nabla_V^2\Gc(x,y) \nonumber\\
D_C\frac{\partial \Gc(x,y,0)}{\partial z} &= k_{on}\int_0^{L_I}\Gc(x,y,z)dz - k_{off}\Gm(x,y)\,.
\end{align}
We have introduced the 2 and 3D surface or volume Laplace operators, $\nabla_S^2$ and $\nabla_V^2$, respectively. If we assume that the concentration in the cytosol does not vary much on a length scale of $L_I$, we can reduce these equations to the local model given by equation \eqref{eq:membraneCytosolBasicEquation}.

\subsection{Binding-Unbinding Equilibrium}\label{sec:App:BindingEquilibrium}
We now consider the equilibrium condition between the binding and unbinding to and from the membrane, which is obtained from equation \eqref{eq:3DmodelEquations} by equating $L_I k_{on}{\Gc}= k_{off}\Gm$. The rates of binding and unbinding are set by $k_{off}$ and $\frac{2k_{on}L_I}{R}$, respectively, where $R$ is the cylinder radius. If $f$ denotes the fraction of membrane-bound to total concentration,
\begin{align}\label{eq:fractionBoundDefinition}
f = \frac{\int_S \Gm dS}{\int_V \Gc dV + \int_S \Gm dS}
\end{align}
then, in a homogeneous equilibrium, we get
\begin{equation}\label{eq:App:fractionBoundGeneral}
 f = \frac{k_{on}}{k_{on}+k_{off}\frac{V}{L_I S}},
\end{equation}
as argued in the main text in equation \eqref{eq:fractionBoundGeneral}.

\section{Dimensional Reduction of the 3D Model}\label{sec:App:DimensionalReduction}
We are now deriving the dimensional reduction of the 3D membrane binding/unbinding model \eqref{eq:membraneCytosolBasicEquation}.

\subsection{1D Reduction}\label{sec:App:oneDimensionalReduction}

We consider a cylindrical cell where the height of the cell is $L$ and the radius is $R$, such that $L\gg R$. We choose cylindrical coordinates $(r,\phi,z)$ such that $r\in[0,R]$, $z\in[0,L]$ and cylindrical symmetry, so our fields $\Gm$ and $\Gc$ do not depend on $\phi\in[0,2\pi]$. As $\Gm$ is the concentration of membrane bound molecules, it is only defined at the boundary of the cell located at $r=R$ and $z=0,L$. Furthermore, the dependence of $\Gc$ on $r$ is weak, relative to the dependence on $z$, due to fast radial diffusion due to $L\gg R$. We completely neglect the $r$ dependence of $\Gm$ as this would only matter at $x=0,L$, and would be weak, similar to the weakness of the $r$ dependence of $\Gc$.

The PDEs in cylindrical coordinates then become
\begin{align}\label{eq:PDEsCyclindricalCoordinates}
 \frac{\partial \Gm(z,t)}{\partial t} &= D_M \partial_z^2 \Gm(z,t) + k_{on}L_I \Gc(R,z,t) - k_{off}\Gm(z,t)\nonumber\\
\frac{\partial \Gc(r,z,t)}{\partial t} &= D_C \left(\partial_z^2+\frac{1}{r}\partial_r r\partial_r \right)\Gc(r,z,t) \nonumber\\
D_C\partial_r \Gc(R,z,t) &= -k_{on}L_I \Gc(R,z,t) + k_{off}\Gm(z,t),\nonumber\\
D_C\partial_z \Gc(r,0,t) &= k_{on}L_I \Gc(r,0,t) - k_{off}\Gm(0,t),\nonumber\\
D_C\partial_z \Gc(r,L,t) &= -k_{on}L_I \Gc(r,L,t) + k_{off}\Gm(L,t).
\end{align}
Then, let us define the following 1D densities
\begin{align}
 {\Gct}(z,t) &= \int_0^R\int_0^{2\pi} \Gc(r,z,t) r d\phi dr\nonumber\\
&=2\pi\int_0^R \Gc(r,z,t)r dr,\nonumber\\
 {\Gmt}(z,t) &= \int_0^{2\pi}\Gm(z,t)Rd\phi = 2\pi R \Gm(z,t).
\end{align}
As $R$ is assumed to be small such that diffusion in the radial direction is faster than other timescales in the problem, radial diffusion will quickly homogenize $\Gc$ in the radial direction even in the case when the initial conditions have a strong $r$ dependence. Hence, ignoring potential fast transient changes of $\Gc$, we consider $\Gc$ to have a weak dependence on $r$ and expand ${\Gct}(z,t)$ to get
\begin{align}
 {\Gct}(z,t) &= 2\pi\int_0^R \left(\Gc(R,z,t)+\partial_r\Gc(R,z,t)(r-R)+\dots\right) r dr\nonumber\\
&\approx 2\pi\left(\Gc(R,z,t) \frac{R^2}{2} - \frac{R^3}{6}\frac{1}{D_C}(k_{off}\Gm(z,t)-k_{on}L_I \Gc(R,z,t))\right)\nonumber\\
&= \pi R^2\left(1+\frac{RL_I}{3D_C}k_{on}\right)\Gc(R,z,t) - \frac{k_{off}R^2}{6D_C}{\Gmt}(z,t).
\end{align}
Here, we have expanded about $r=R$ and used the boundary condition. If we further assume that the radius $R$ and the interaction range $L_I$ are small such that the equation \eqref{eq:condition1Dmodel1} holds, then we can approximate further (see section \ref{sec:App:parameters} for a discussion of those parameters)
\begin{align}
 {\Gct}(z,t) &\approx \pi R^2 \Gc(R,z,t).
\end{align}
The 1D densities follow the differential equations
\begin{align}
 \frac{\partial{\Gmt}(z,t)}{\partial t} &= D_M\partial_z^2 {\Gmt}(z,t) + 2\pi Rk_{on}L_I \Gc(R,z,t)) - k_{off}{\Gmt}(z,t),\nonumber\\
&= D_M\partial_z^2 {\Gmt}(z,t) + 2k_{on}\frac{L_I}{R} {\Gct(z,t)} - k_{off}{\Gmt}(z,t),\nonumber\\
 \frac{\partial {\Gct}(z,t)}{\partial t} &= D_C\left(\partial_z^2 {\Gct}(z,t) +\int_0^R\left(\frac{1}{r}\partial_rr\partial_r\Gc(r,z,t)\right)2\pi r dr\right),\nonumber\\
&=D_C\left(\partial_z^2 {\Gct}(z,t) +2\pi R\partial_r \Gc(R,z,t)\right),\nonumber\\
&=D_C\left(\partial_z^2 {\Gct}(z,t) +2\pi R\frac{1}{D_C}\left( k_{off}m(z,t)-k_{on}L_I \Gc(z,t) \right)\right),\nonumber\\
&=D_C\partial_z^2 {\Gct}(z,t) + \left( k_{off}{\Gmt}(z,t)-2k_{on}\frac{L_I}{R} {\Gct}(z,t) \right),\nonumber\\
\end{align}
which were given in the main text in equation \eqref{eq:cylinderReductionTo1D}. Mass conservation is ensured by accompanying those equations by Neumann no-flux boundary conditions. Note that while these equations are perfectly 1D PDEs, with the spatial domain defined by the length of the cylinder, the cylinder radius is still felt in the sense that the membrane-binding rate $k_{on}$ is effectively renormalized by the inverse of the cylinder radius $R$.

\subsubsection{Equilibrium in One Spatial Dimensional}

We consider the binding-unbinding equilibrium condition for the cylindrical cell in the 1D limit, as done in section \ref{sec:App:BindingEquilibrium} for a general 3D cell. We obtain for the fraction $f$ of membrane-bound molecules
\begin{align}\label{eq:App:fractionBound1Dcylinder}
f = \frac{k_{on}}{k_{on}+k_{off}\frac{R}{2 L_I}} = \frac{k_{on}}{k_{on}+k_{off}\sqrt{\frac{V}{4\pi L_I^2L}}},
\end{align}
as argued in equation \eqref{eq:fractionBound1Dcylinder}. Notice that the dependence of $f$ on $V$ and $L$ is different to the one found in \cite{holmes2012modelling}, and we have checked that this difference is not because of the use of rectangular, rather than cylindrical cells. The binding/unbinding equilibrium and simultaneous 1D limit are valid if the parameters satisfy the constraints
\begin{align}\label{eq:condition1Dmodel2}
\frac{D_C}{R^2}\gg k_{off} &\gg \frac{D_C}{L^2},\nonumber\\
\frac{D_C}{R^2}\gg \frac{2k_{on}L_I}{R} &\gg \frac{D_C}{L^2}.
\end{align}
These constraints ensure diffusion in the radial direction occurs on the fastest timescale to ensure there is only a weak radial dependence, and a 1D limit is justified. On the other hand, binding and unbinding are faster than diffusion along $L$ so the binding/unbinding equilibrium is justified. We can then introduce an effective diffusion coefficient $D_{MC}= f D_M + (1-f)D_C$, similar as in \cite{dawes2007phosphoinositides,holmes2012modelling}, so that the total concentration $G^{tot}(z,t) = {\Gct}(z,t) + {\Gmt}(z,t)$ simply evolves by the standard 1D diffusion equation
\begin{align}\label{eq:effectiveDiffusionInactiveGTPase1D}
\frac{\partial G^{tot}(z,t)}{\partial t} &= D_{MC}\frac{\partial^2G^{tot}(z,t)}{\partial z^2}.
\end{align}
We have thus reduced the original system of two coupled PDEs in three spatial dimensions to a single PDE in one spatial dimension. Note that other potential reactions in the system would need to be modified by $f$ accordingly, and, if they are present, the timescales of associated with those reactions need to be compared to the timescales of radial diffusion and binding/unbinding to justify the reduction of the complete model with reactions to lower dimensions.

\subsection{2D Reduction}\label{sec:App:twoDimensionalReduction}

Let us consider a flat cell, which, for simplicity, we take to be a disk of radius $R$ and height $h$. Hence, it is natural to choose cylindrical coordinates, and the full model is described by \eqref{eq:PDEsCyclindricalCoordinates}, with $L$ replaced by $h$. We now consider the limit $h\ll R$. Then, we can rewrite the $z$-dependence of the Laplacian as
\begin{align}
 D_C\frac{\partial^2 \Gc(r,\phi,z,t)}{\partial_z^2} &\approx D_c\frac{\partial_z\Gc(r,\phi,h)-\partial_z\Gc(r,\phi,0)}{h},\nonumber\\
&= \frac{k_{off}}{h}\left(\Gm(r,h,t)+\Gm(r,0,t)\right) \nonumber\\
&-\frac{k_{on}L_I}{h}\left(\Gc(r,\phi,h,t)+\Gc(r,\phi,0,t)\right).
\end{align}
If the concentrations only depend very weakly on $z$, then we can simplify the system by introducing 
\begin{align}
 {\Gch}(r,\phi,t) &= \int_0^h \Gc(r,\phi,z,t) d z dr = h \Gc(r,\phi,t),\nonumber\\
{\Gmh}(r,\phi,t) &= 2 \Gm(r,\phi,t),
\end{align}
and get 
\begin{align}
\frac{\partial {\Gmh}(r,\phi,t)}{\partial t} &= D_M \nabla^2_P {\Gmh}(r,\phi,t) + 2k_{on}\frac{L_I}{h} {\Gch}(r,\phi,t) - k_{off}{\Gmh}(r,\phi,t),\nonumber\\
\frac{\partial {\Gch}(r,\phi,t)}{\partial t} &= D_C \nabla^2_P {\Gch}(r,\phi,t) - 2 k_{on}\frac{L_I}{h} {\Gch}(r,\phi,t) + k_{off}{\Gmh}(r,\phi,t).
\end{align}
Here, $\nabla^2_P$ denotes the conventional 2D Laplace operator in polar coordinates. Similarly to the 1D case, the scaling of the parameters is different. Similarly to the 1D case, we find that the parameters of the reduced geometry, in this case, the cylinder height $h$, renormalize the effective membrane-binding coefficient. As before, we consider the steady-state solution where $f$ denotes the fraction of membrane-bound molecules. Hence, for the oblate cylinder we get
\begin{equation}
 f = \frac{k_{on}}{k_{on}+k_{off}\frac{h}{2L_I}}.
\end{equation}
%

\section{3D Polarization Pathway}\label{sec:App:3DSimulations}
In this section, we discuss in detail how our 3D pathway model discussed in section \ref{sec:pathway}, is obtained and relates to the 1D model discussed in \cite{holmes2012modelling}.
\subsection{Model Setup}
The 1D model of \cite{holmes2012modelling} was motivated by experiments where cells were constrained in effective 1D geometries. It was assumed that the approximate 3D geometry is rectangular with length scales $L\gg w > d$, with an initial length of $L=20\mu m$. The volume of a cell in the experimental paper \cite{lin2012synthetic}, which uses the model of \cite{holmes2012modelling}, was given as approximately $V=800\mu m^3$. The supplementary information of \cite{lin2012synthetic} mentions $d=0.2\mu m$, which seems a bit small for a real cell and would also imply that $w=200\mu m$ at the given volume. We hence compare to a cell with base measure $L,w,d=20,8,5\mu m$. In \cite{lin2012synthetic}, the cells were about $80\mu m$ long (see e.g. figure S4 in \cite{lin2012synthetic}), which would, at the same volume, be compatible with $L,w,d=80,5,2\mu m$. Such cell seems also compatible with the 1D limit as described in the section \ref{sec:oneDimensionalReduction} of the main text. In \cite{holmes2012modelling,lin2012synthetic} the change in cell length was taken into account by changing the fraction of membrane bound to unbound inactive molecules via an equation similar, but slightly different, to our equation \eqref{eq:fractionBound1Dcylinder}. The 3D model will automatically take into account any geometry change. The basic equations for the evolution of the three GTPases now takes into account that the each GTPase can exist in an active, membrane bound form $\Gma$, a membrane bound, inactive form $\Gmi$ and a form $\Gc$ which diffuses inactively through the cytosol. These follow the principal scheme 
\begin{align}\label{eq:GTPaseequations}
\frac{\partial \Gma}{\partial t} &= D_M \nabla^2_S \Gma + I_G \Gmi - \delta_G \Gma, \nonumber\\
\frac{\partial \Gmi}{\partial t} &= D_M \nabla^2_S \Gmi - I_G \Gmi + \delta_G \Gma + k_{on}L_I \Gc - k_{off}\Gmi,\nonumber\\
\frac{\partial \Gc}{\partial t} &= D_C \nabla^2_V \Gc, \nonumber\\
 -D_C e_n \nabla_V \Gc &= k_{on}L_I \Gc - k_{off}\Gmi.
\end{align}
Here, $I_G$ represents the activation, and $\delta_G$ the deactivation rate, whereas $k_{on},k_{off}$ denote the binding and unbinding rates as in the section \ref{sec:MembraneCytosolBinding}. Note that equations \eqref{eq:GTPaseequations} are slightly different from the equations given in the appendix of \cite{holmes2012modelling}, which were used to motivate the 1D model from a 3D perspective. 

\begin{figure}

 \includegraphics[width=0.2\linewidth]{Stimulus_arrow_right_top}
 \includegraphics[width=0.2\linewidth]{Stimulus_arrow_left_top}
 \includegraphics[width=0.2\linewidth]{Stimulus_arrow_left_top}
 \includegraphics[width=0.2\linewidth]{Stimulus_arrow_left_top}

\subfloat[t=100s]
{
 \includegraphics[width=0.2\linewidth]{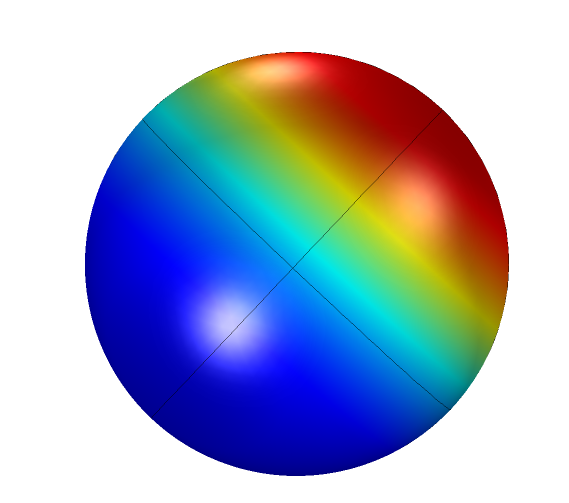}
}
\subfloat[t=140s]
{
 \includegraphics[width=0.2\linewidth]{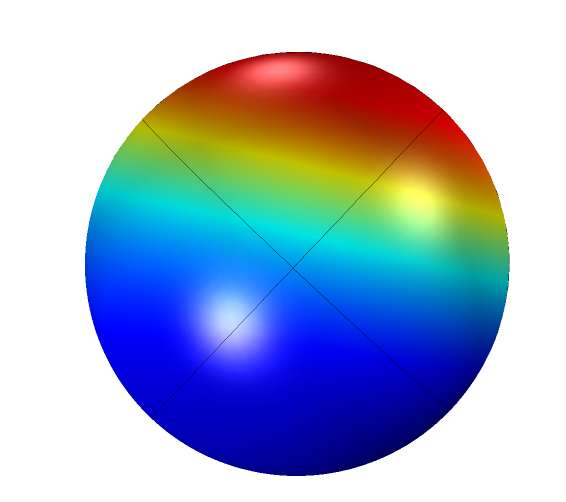}
}
\subfloat[t=180s]
{
 \includegraphics[width=0.2\linewidth]{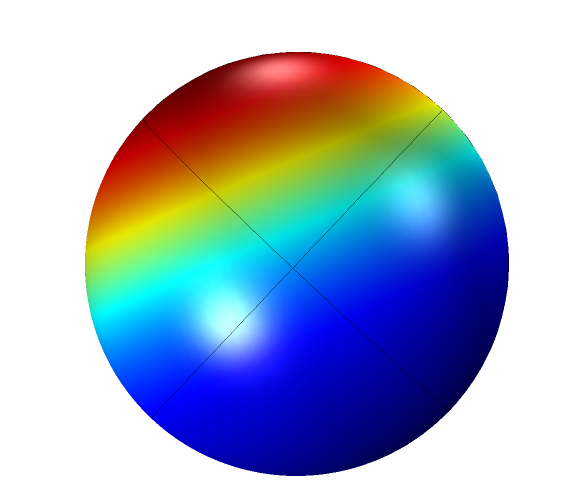}
}
\subfloat[t=300s]
{
 \includegraphics[width=0.2\linewidth]{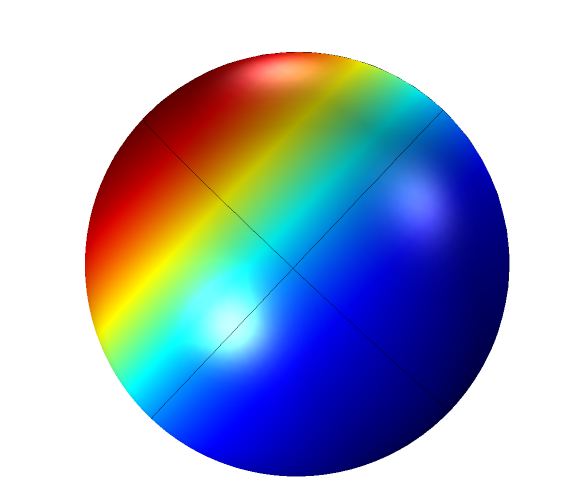}
}
\includegraphics[width=0.13\linewidth]{rotation_legend}
\\
\subfloat[t=100s]
{
 \includegraphics[width=0.2\linewidth]{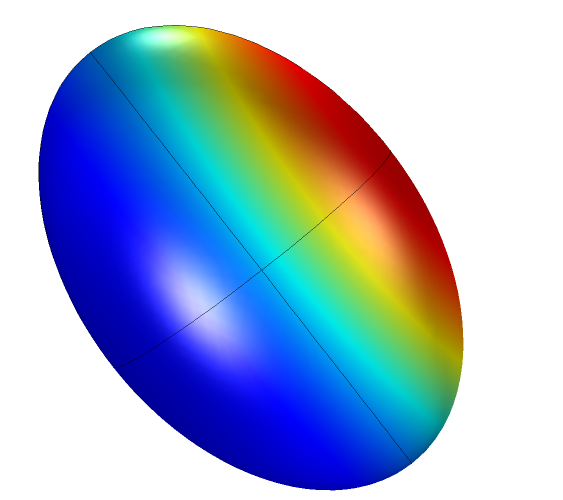}
}
\subfloat[t=140s]
{
 \includegraphics[width=0.2\linewidth]{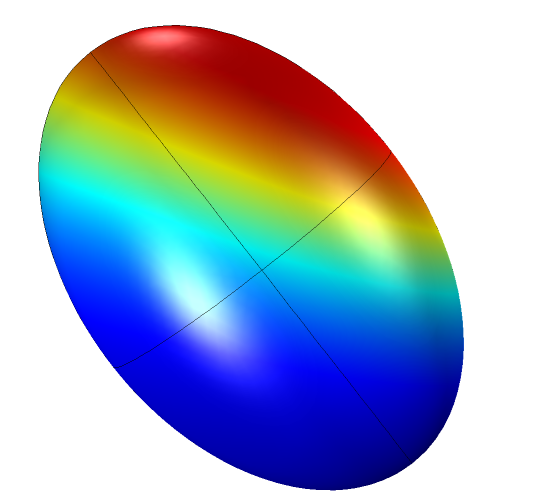}
}
\subfloat[t=180s]
{
 \includegraphics[width=0.2\linewidth]{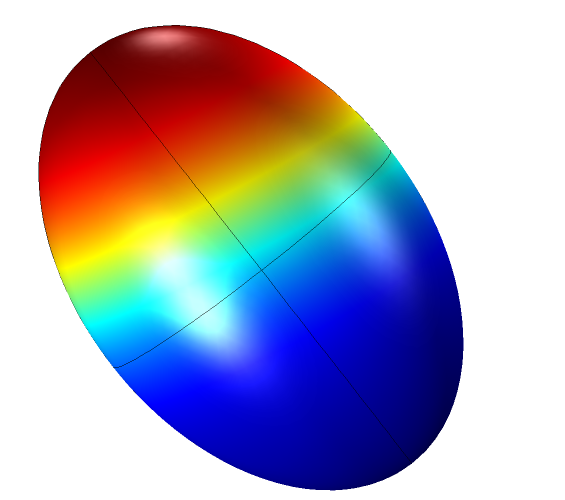}
}
\subfloat[t=300s]
{
 \includegraphics[width=0.2\linewidth]{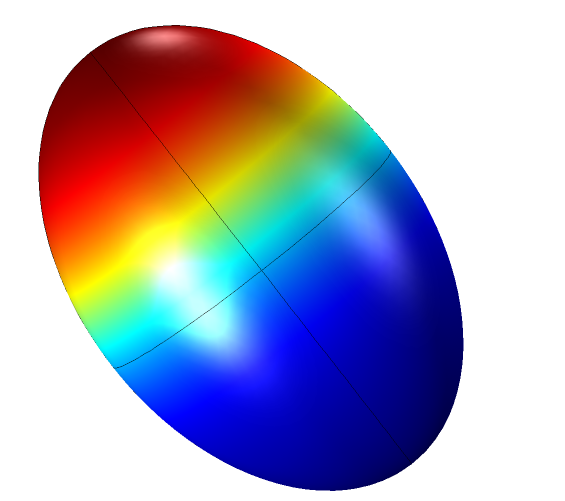}
}
\includegraphics[width=0.13\linewidth]{rotation_legend}
\\
\subfloat[t=100s]
{
 \includegraphics[width=0.2\linewidth]{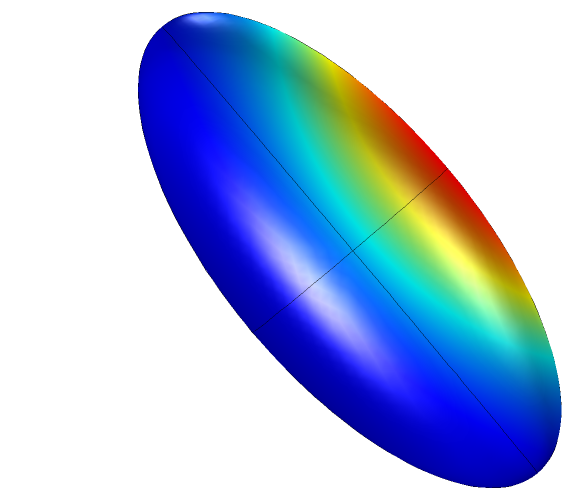}
}
\subfloat[t=140s]
{
 \includegraphics[width=0.2\linewidth]{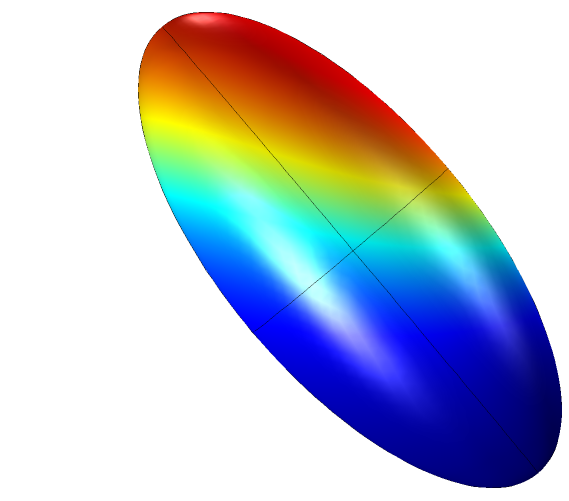}
}
\subfloat[t=180s]
{
 \includegraphics[width=0.2\linewidth]{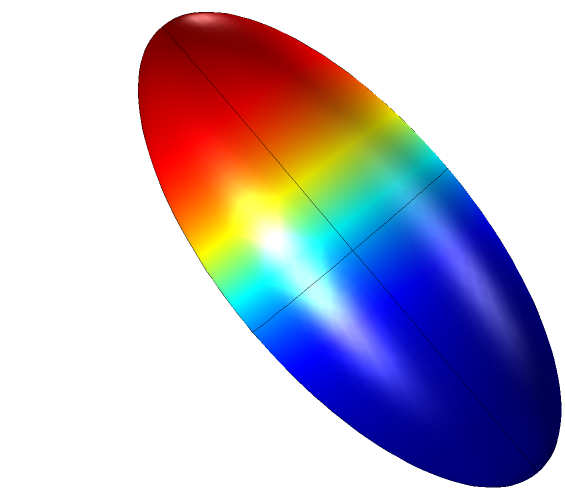}
}
\subfloat[t=300s]
{
 \includegraphics[width=0.2\linewidth]{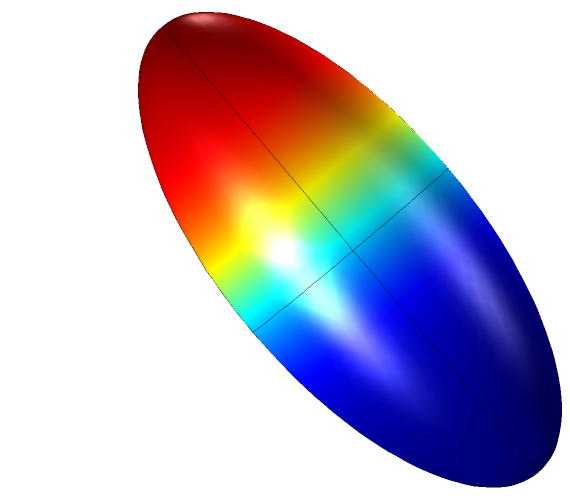}
}
\includegraphics[width=0.13\linewidth]{rotation_legend}
\caption{\label{fig:RepolarizationOfEllipsoidsFromShortAxisFullRotation} As in figures \ref{fig:RepolarizationOfEllipsoids}, \ref{fig:RepolarizationOfEllipsoidsFromShortAxis}, active Rac on the membrane is shown for different times and cells of different shapes, but here, the Rac activation rate in the first $100 s$ increases linearly along a short axis of the ellipsoid (from lower left corner to upper right corner), and from then on, it increases linearly along the long axis of the ellipsoids (from the lower right corner to the upper left corner). In all cases, the volume of the ellipsoid cells is fixed as $V=800\mu m^3$, the main axis is $11.5\mu m$ (spherical, (a)-(d)), $15\mu m$ ((e)-(h)) and $20\mu m$ ((i)-(l)), and the other two axes are of the same length. Cells of all shapes are able to adapt to their new stimulus direction, with the elongated cells being slightly faster.}
\end{figure}

We note that to account for the proper localization of the membrane bound and unbound species $\Gma,\Gmi$ and $\Gc$, we measure $\Gc$ in Molar, but $\Gma,\Gmi$ in $\frac{mol}{m^2}$. Whereas this latter measure is not often chosen in experiments, as usually total cell concentrations are measured, this is nevertheless the physically more meaningful measure, as $\Gma,\Gmi$ denote number molecules per two dimensional membrane area, and this choice ensures that our equations and the dimensional reductions have the correct units. To compare with \cite{holmes2012modelling}, we will hence multiply the concentrations of active GTPases with $\frac{V_0}{S_0}=\frac{800\mu m^3}{600\mu m^2}=4/3 \mu m$, the fraction of volume to surface area for the above mentioned rectangular cell of basic length $L,w,d=20,8,5\mu m$. With this setup, it is straight-forward to generalize the 1D model from \cite{holmes2012modelling}, summarized in figure \ref{fig:pathway}, in our 3D context, and one obtains equations \eqref{eq:3DmodelEquations}. All coefficients apart from the membrane binding and unbinding rates are taken from \cite{holmes2012modelling,lin2012synthetic}, but those which multiply a membrane density are multiplied by the factor $\frac{V_0}{S_0}$. Furthermore, to compare the activation rates given in \cite{holmes2012modelling,lin2012synthetic} with ours we need to divide them with the fraction of bound to total inactive molecules $f$, as we separately consider bound and unbound inactive GTPases. 
We now summarize the use of the spatially dependent Rac stimulus $S_{Rac}$ which appears in the activation rate for Rac, equation \eqref{eq:activationFunctions} in most simulations shown in section \ref{sec:3DSimulations}. If the x-axis denotes the direction of the stimulus, then typically we assumed a linear stimulus of strength $S_{Rac} = 0.5 I_{R1}\frac{x}{20\mu m}$, where $I_{R1}$ is the baseline Rac activation rate as given in Table \ref{tab:Parameters}. The baseline length of $20\mu m$ is chosen to match results from earlier works, as described above. In figures \ref{fig:RepolarizationOfEllipsoids}, \ref{fig:RepolarizationOfEllipsoidsFromShortAxis} and \ref{fig:RepolarizationOfEllipsoidsFromShortAxisFullRotation} such gradient was used but then rotated towards the indicated axis at $t=100s$.

\subsection{Parameter Estimation}\label{sec:App:parameters}
We now investigate when the use of the 1D model is justified for the case where the molecule is a small GTPase. We have $D_C=100 \mu m^2/s$ \cite{dawes2007phosphoinositides,lin2012synthetic}. Furthermore, we estimate that the interaction range of the binding reaction, $L_I$, approximately corresponds to the size of the molecules. We have a molecular weight of the small GTPases of about $21 kDa$. Exact size determination of proteins is tricky \cite{erickson2009size}, but here we only need a rough estimate, which gives that we have a volume of $V=\frac{21 * 1.6 *10^{-27}kg}{1kg/l}=3*10^{-26}m^3=30nm^3$. Hence, the interaction length scale is on the order of a few nanometers. We put $L_I=2nm$, and this estimate is similar to stimations made in similar contexts \cite{mclaughlin1995myristoyl}. Furthermore, we can safely assume that $R\gg L_I$ for realistic cell geometries. Then, unless ${k_{on}}\gg k_{off}$, of the two requirements $\frac{RL_I{k_{on}}}{3D_C} \ll 1, \frac{k_{off}R^2}{6D_C}\ll 1$, the first one automatically holds provided the second one does. From \cite{moissoglu2006vivo} we can estimate that $k_{off}$ should be faster than $k_{off}=0.06 s^{-1}$, as the combined deactivation/unbinding rate (this combined rate is denoted $k_{off}$ in \cite{moissoglu2006vivo}) is of this magnitude. However, the actual binding and unbinding rates are influenced by the presence of other regulators such as GDI molecules \cite{dovas2005rhogdi} and might be different for GTP and GDP bound GTPases, and is hence also influenced by the presence of GEFs and GAPs. Here, we focus on rough estimates and use the above numbers to derive a limit for the radial length scale of
\begin{equation}
 R \ll\sqrt{\frac{6D_C}{k_{off}}} \leq \sqrt{\frac{600}{0.06}}\mu m = 100\mu m.
\end{equation}
As long as $k_{off}$ is not too large this condition is satisfied for realistic cell dimensions. However, if $k_{off}$ should be significantly larger than estimated above then this limit might be hard to satisfy.

Now we look at the steady-state assumption between bound and unbound GTPase. For this, we have
\begin{equation}
 L \gg\sqrt{\frac{D_C}{k_{off}}}.
\end{equation}
With $k_{off}=0.06 s^{-1}$, we get $L\gg 30\mu m$ would safely satisfy this constraint. However, most likely $k_{off}$ is significantly larger so the steady-state assumption is most likely valid for shorter cells as well.

All other parameters used in equations \eqref{eq:3DmodelEquations} are summarized in Table \ref{tab:Parameters}.
\begin{table}
\begin{center}
\begin{tabular}{ccccc}
\small{Dimensional parameter} & \small{Estimate} \\
$D_C$ & 100 $\mu $m$^2$ s$^{-1}$ \\
$D_M$ & $1 \mu $m$^2$ s$^{-1}$ \\ 
$D_P$ & $5\mu $m$^2$ s$^{-1}$ \\ 
$\text{Rac}_{tot,2}$ & 10nmol m$^{-2}$ \\ 
$\text{Rho}_{tot,2}$ & 4nmol m$^{-2}$ \\
$\text{Cdc}_{tot,2}$ & 3.4nmol m$^{-2}$ \\
$P_{3b}$ & 0.2nmol m$^{-2}$ \\
$\delta_R$ & $1$s$^{-1}$ \\ 
$\delta_{\rho}$ & $1$s$^{-1}$ \\ 
$\delta_C$ & $1$s$^{-1}$ \\ 
$\delta_{P_1}$ & $0.21$s$^{-1}$ \\ 
$k_{21}$ & $0.021$s$^{-1}$ \\ 
$k_{\text{max}}$ & $2.8$s$^{-1}$ \\ 
$k_{P_2}$ & $2.1$s$^{-1}$ \\ 
$\alpha$ & $1.3$s$^{-1}$ \\ 
$\mu_P$ & $0.011$s$^{-1}$ \\ 
$G$ & $0.03$s$^{-1}$ \\ 
$I_{R1}$ & $0.4\mu$M s$^{-1}$ \\ 
$I_{R2}$ & $0.4\mu$M s$^{-1}$ \\ 
$I_{\text{Rho}}$ & $13.2\mu$Ms$^{-1}$ \\ 
$I_{\text{Cdc}}$ & $5.9\mu$Ms$^{-1}$ \\ 
$I_{P_1}$ & $14$nmol m$^{-2}$ s$^{-1}$ \\ 
$a_1$ & $1.7$nmol m$^{-2}$ \\ 
$a_2$ & $1.3$nmol m$^{-2}$ \\ 
$k_{\text{PI5K}}$ & $0.084$s$^{-1}$ \\ 
$k_{\text{PI3K}}$ & $0.00072$s$^{-1}$ \\ 
$k_{\text{PTEN}}$ & $0.432$s$^{-1}$ \\ 
$f_1$ & $1$ \\
$k_{off}$ & $10$s$^{-1}$ \\
$k_{on}L_I$ & $13.3\mu $m s$^{-1}$\\
\hline
\end{tabular}
\end{center}
\caption{\label{tab:Parameters}
Parameters of the cell polarization model equation \eqref{eq:3DmodelEquations}, inferred for a cell with dimensions $L,w,d=20,8,5\mu m$, so all values which multiply membrane concentrations are rescaled by the factor $\frac{V_0}{S_0} = \frac{4}{3}\mu m$. This means $1\mu M \frac{V_0}{S_0} = \frac{4}{3}\frac{nmol}{m^2}$. Furthermore, $\alpha$, $I_{R1}$, $I_{R2}$ $I_{\text{Rho}}$ and $I_{\text{Cdc}}$ are multiplied by the baseline fraction of bound inactive molecules $f$. We have also rounded the parameters as appropriate.}
\end{table}
Lacking accurate measurements of $k_{off}$, it is commonly believed that $k_{off}$ is much faster larger than the deactivation rate \cite{moissoglu2006vivo,holmes2012modelling}. As the deactivation rates were estimated in \cite{holmes2012modelling} to be $1 s^{-1}$, we take $k_{off}=10 s^{-1}$. While the total unbinding/deactivation rate was estimated in \cite{moissoglu2006vivo} to be much smaller than $1 s^{-1}$, we stick to those values here as we first would like to compare our 3D model to the 1D model of \cite{holmes2012modelling}. In the subsection \ref{sec:parameterStudy} in the main text we study the influence of varying $k_{off}$ on the polarization behavior of the cell. We also need to determine the combination of parameters $k_{on}L_I$. From the estimates of $D_C, D_M$, combined with the estimate that at baseline length of $L=20\mu m$ the diffusion coefficient for total inactive GTPases (bound and unbound) is $50\mu m^2/s$, we get that about half of the inactive GTPases molecules are typically membrane bound. This equilibrium value can then be used to deduce $k_{on}L_I$ via equation \eqref{eq:fractionBoundGeneral}.

\subsection{Implementation of the 3D Model}
All simulations of the 3D model were performed in COMSOL Multiphysics 5.1 (COMSOL, Inc, Burlington, MA) using the General PDE model framework. We remark that the default solver occasionally produced too large time steps, requiring us to manually limit the maximal time step depending on the model parameters. It is also necessary to choose a fine mesh for good spatial resolution in several cases, for instance, when the cytosolic species vary sharply at the membrane. In most cases, the predefined Mesh Element settings 'Finer' or 'Extra Fine' were sufficient to ensure spatial convergence.

\newpage
\subsection*{}%
\begin{figure}[ht]
\begin{center}
\subfloat[$L=20\mu m$]
{
 \includegraphics[width=0.55\linewidth]{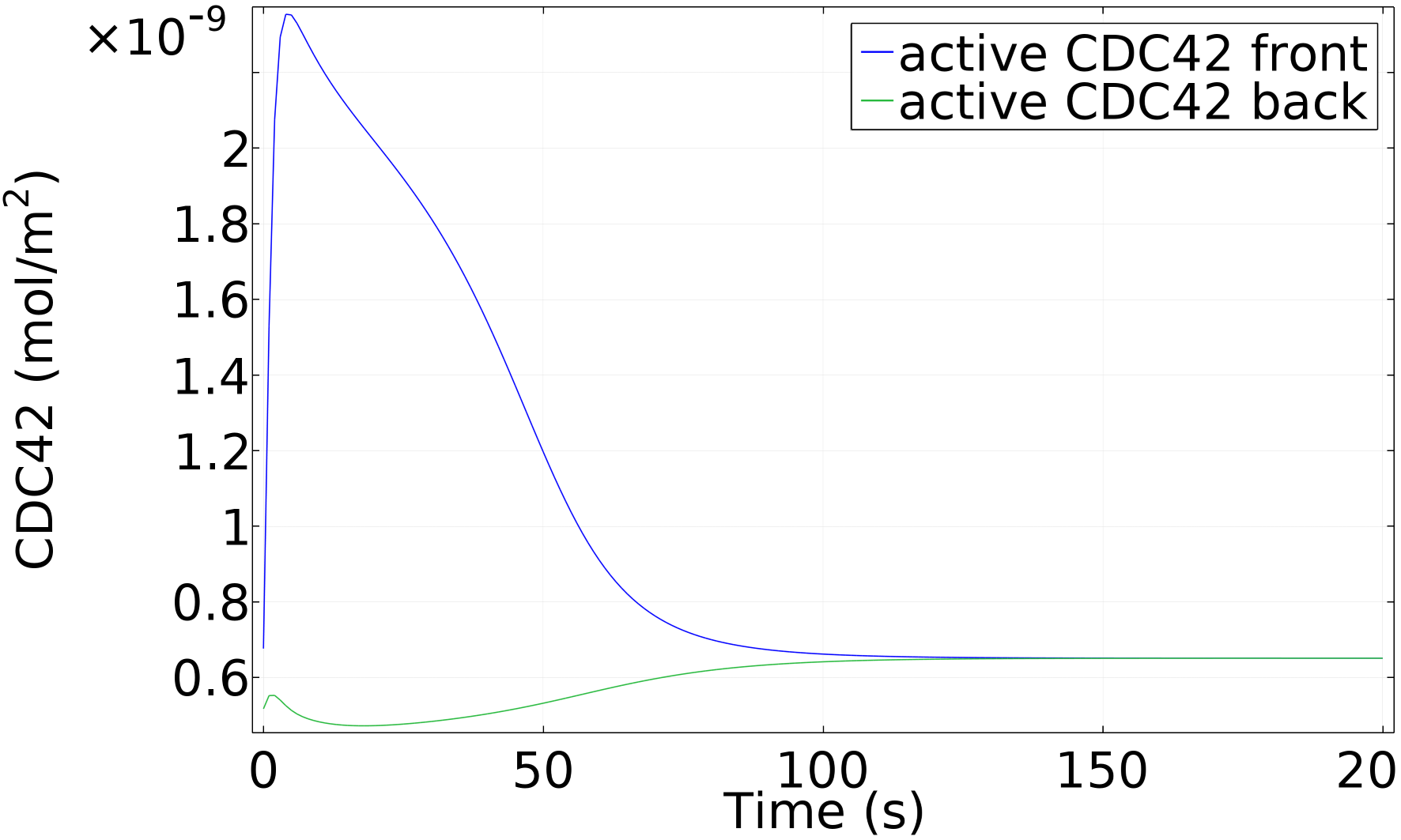}
}
\end{center}
\begin{center}
\subfloat[$L=40\mu m$]
{
 \includegraphics[width=0.55\linewidth]{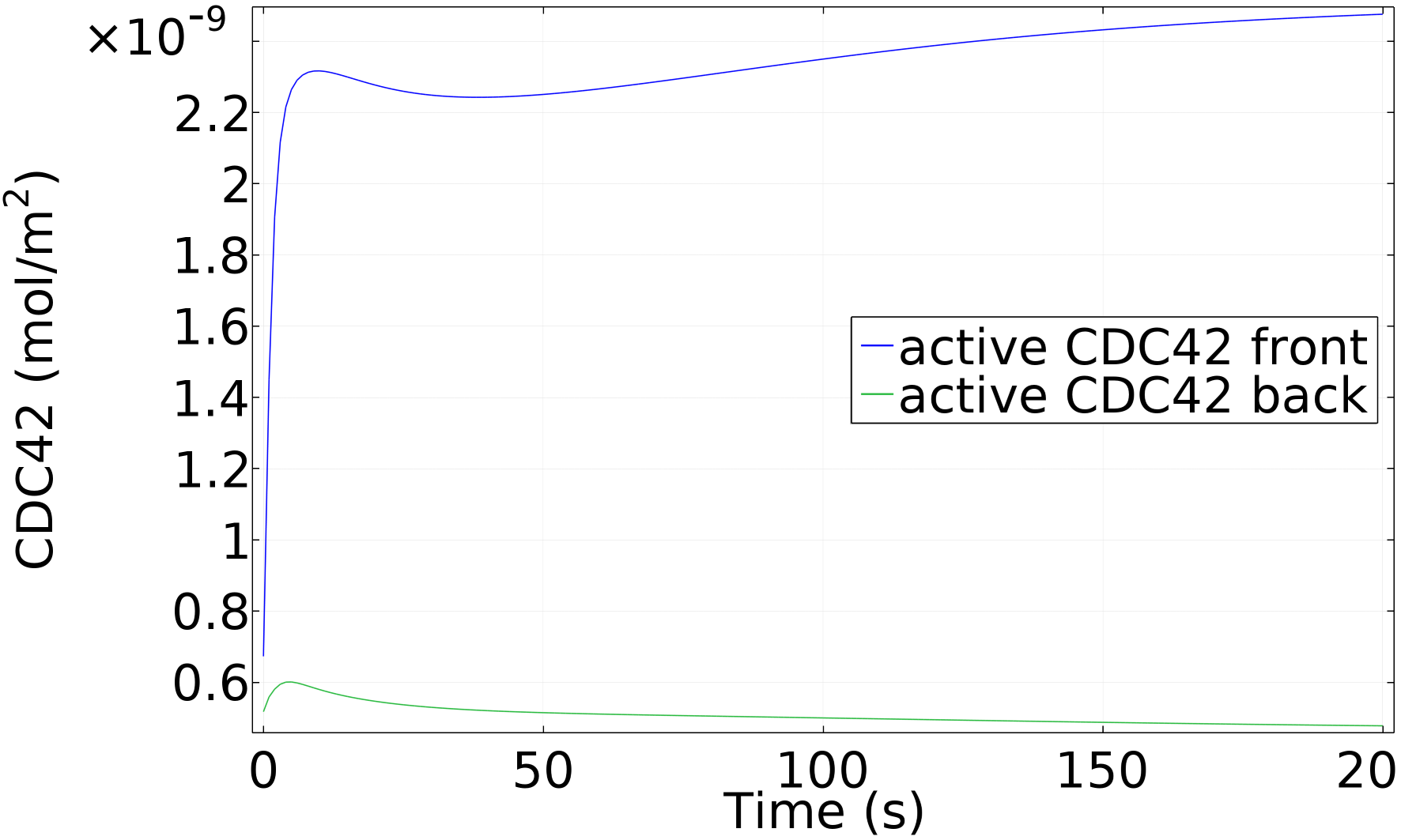}
}
\end{center}
\begin{center}
\subfloat[$L=80\mu  m$]
{
 \includegraphics[width=0.55\linewidth]{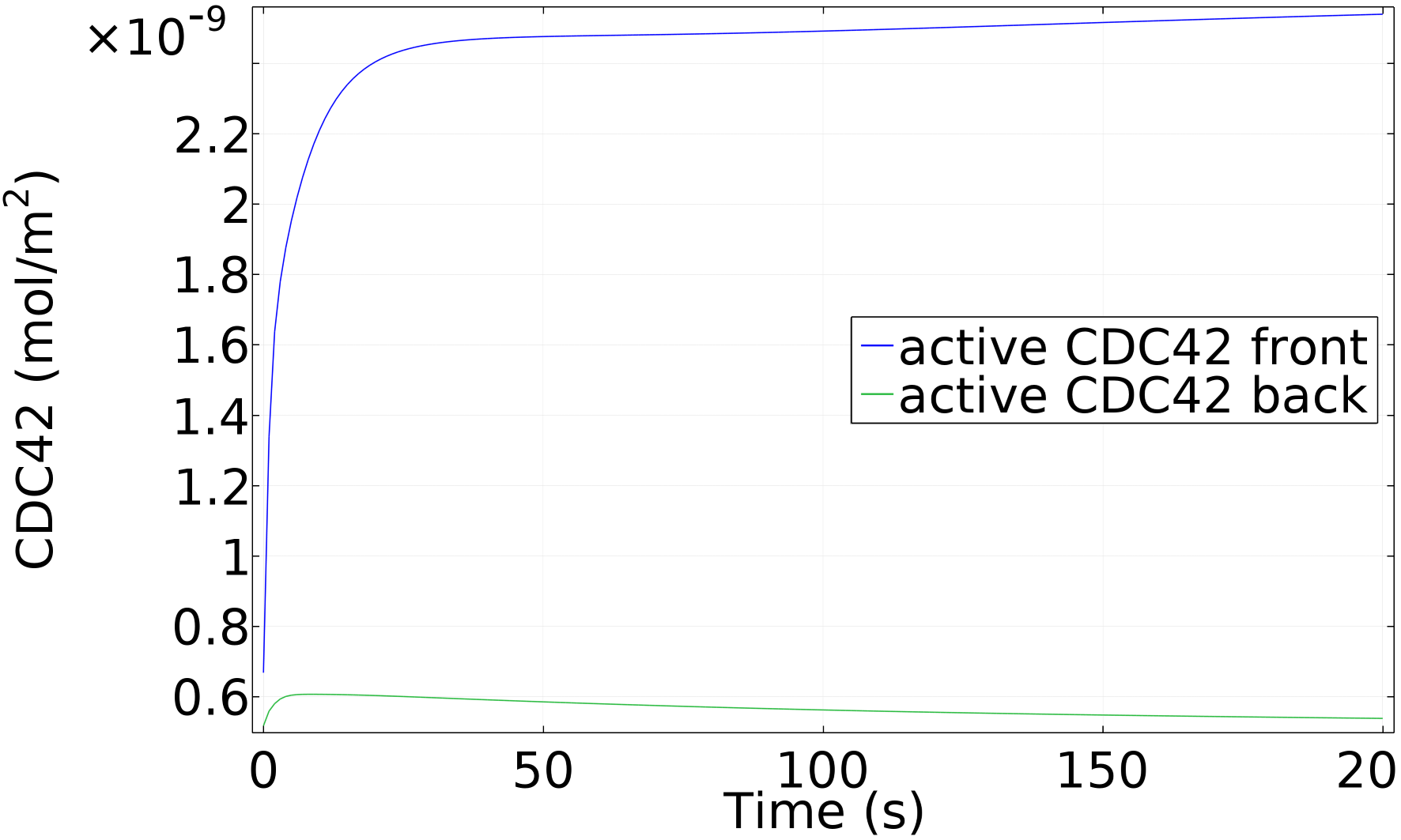}
}
\end{center}
\caption{\label{fig:App:PolarizationOfRectangularDifferentLengthInTime}
 The time evolution of concentration of active Cdc42 in time at the front and back of the rectangular cells as shown in figures \ref{fig:NonPolarizedRectangular} and \ref{fig:PolarizationOfRectangularDifferentLength}.
}
\end{figure}

\bibliography{migration}{}
\bibliographystyle{unsrt}
%
%

\end{document}